\documentclass[11pt,a4paper]{article}
\pdfoutput=1 

\usepackage{jheppub} 

\usepackage[T1]{fontenc} 
\usepackage[utf8]{inputenc}

\usepackage{subfigure}
\usepackage{array,multirow}
\usepackage{etoolbox}
\usepackage{booktabs}
\usepackage{float}
\usepackage{scalerel}
\usepackage{epsfig,color,wrapfig}
\usepackage{amsmath}
\usepackage{graphicx}
\usepackage{amssymb}
\usepackage{slashbox}
\usepackage{comment}
\usepackage{bm}
\usepackage[toc]{appendix}
\usepackage{slashed}
\usepackage{caption}
\usepackage{dcolumn,multirow}
\usepackage[normalem]{ulem}

\usepackage{url}
\usepackage{adjustbox}

{\newcommand{\lsim}{\mbox{\raisebox{-.6ex}{~$\stackrel{<}{\sim}$~}}}
	{

		\def\bdstlnu{$B \to D^* \ell \nu_{\ell}$ }
		\def\bdlnu{$B \to D \ell \nu_{\ell}$ }
		\def\bdsttaunu{$B \to D^* \tau \nu_{\tau}$ }
		\def\bdtaunu{$B \to D \tau \nu_{\tau}$ }
		\def\errvec#1{$\left(\begin{smallmatrix} #1 \end{smallmatrix}\right)$}
		
		\newcommand{\bmt}{\begin{pmatrix}}
			\newcommand{\emt}{\end{pmatrix}}
		\newcommand{\ba}{\begin{array}{c}}
			\newcommand{\ea}{\end{array}}
		\newcommand{\be}{\begin{equation}}
		\newcommand{\ee}{\end{equation}}
		\newcommand{\bea}{\begin{eqnarray}}
		\newcommand{\eea}{\end{eqnarray}}
		\newcommand{\nn}{\nonumber}
		\newcommand{\bi}{\begin{itemize}}
			\newcommand{\ei}{\end{itemize}}
		
		\newcommand{\baz}{\begin{array}{cc}}
			
			\newcommand{\mathsym}[1]{{}}

			\newcommand{\bt}{\begin{tabular}}
				\newcommand{\et}{\end{tabular}}

			\newcommand{\benu}{\begin{enumerate}}
				\newcommand{\eenu}{\end{enumerate}}
			
			\newcommand{\bav}{\begin{array}{cccc}}

\title{\boldmath Updates on extraction of $|V_{cb}|$ and SM prediction of $R(D^{*})$ in $B\to D^{*}\ell\nu_\ell$ decays}

\author[a]{Sneha Jaiswal,}
\author[a]{Soumitra Nandi}
\author[a,b]{and Sunando Kumar Patra}

\affiliation[a]{Indian Institute of Technology, North Guwahati, Guwahati 781039, Assam, India }
\affiliation[b]{Department of Physics, Bangabasi Evening College, 19 Rajkumar Chakraborty Sarani, Kolkata 700009, West Bengal, India }

\emailAdd{sneha.jaiswal@iitg.ac.in}
\emailAdd{soumitra.nandi@iitg.ac.in}
\emailAdd{sunando.patra@gmail.com}

\abstract{We update the standard model (SM) predictions of $R(D^*)$ using the latest results on the decay distributions in $B \to D^* \ell \nu_{\ell}$ ($\ell = \mu, e$) by Belle collaboration, while extracting $|V_{cb}|$ at the same time. Depending on the inputs used in the analysis, we define various fit scenarios. Although the central values of the predicted $R(D^*)$ in all the scenarios have reduced from its earlier predictions in 2017, the results are consistent with each other within the uncertainties. In this analysis, our prediction of $R(D^*)$ is consistent with the respective world average at $\sim 3\sigma$. We have also predicted several angular observables associated with $B \to D^* \tau \nu_{\tau}$ decays. We note that the predicted $F_L(D^*)$ is consistent with the corresponding measurement at 2$\sigma$. Utilizing these new results, we fit the Wilson coefficients appearing beyond the standard model of particle physics (BSM). To see the trend of SM predictions, we have used the recently published preliminary results on the form-factors at non-zero recoil by the lattice groups like Fermilab-MILC and JLQCD and predicted the observables in $B \to D^* \ell \nu_{\ell}$, and $B \to D^* \tau \nu_{\tau}$ decays.}

\begin{document} 
	\maketitle
	\flushbottom

\section{Introduction}
\label{sec:intro}
Precise extraction of the CKM element $|V_{cb}|$ is an important goal of the $B$-physics phenomenology. Inclusive and exclusive tree level semileptonic decays $b\to c\ell\nu_{\ell}$ ($\ell = e, \mu$) are crucial in this regard. Note that the inclusive and exclusive determinations of $|V_{cb}|$ differ by $\sim 3\sigma$ \cite{Alberti:2014yda,Gambino:2016jkc,Amhis:2016xyh}. We will focus on the exclusive determination of $|V_{cb}|$ from $B\to D^*\ell\nu$ decays in this work. Other related observables like $R(D^{(*)})= \frac{Br(B\to D^{(*)}\tau\nu_{\tau})}{Br(B\to D^{(\ast)}\ell\nu_{\ell})}$ are useful for the test of lepton universality. Significant deviations from their respective SM predictions will be a clear signal for the lepton universality violating (LUV) new physics (NP). Precise prediction of these observables, thus, is of utmost importance.

In addition to the considerable improvements in the lattice determination of the form factors in the last decade \cite{Bailey:2014tva, Lattice:2015rga, Na:2015kha}, updated results on the branching fractions $Br(B\to D^{(\ast)}\ell\nu)$ in different $q^2$-bins are now available \cite{Abdesselam:2017kjf}. With these new inputs, several groups had reanalyzed these decay modes using the Boyd-Grinstein-Lebed (BGL) \cite{Boyd:1997kz} and Caprini-Lellouch-Neubert (CLN) \cite{Caprini:1997mu} parametrizations for the form-factors. The latter uses heavy quark effective theory (HQET) relations between the form-factors in which relevant higher-order corrections are missing. The extracted values of $|V_{cb}|$ and $R(D^{(*)})$ have improved over their earlier estimates. For details, see \cite{Bigi:2016mdz,Bigi:2017njr,Grinstein:2017nlq,Bernlochner:2017jka,Bigi:2017jbd,Jaiswal:2017rve} and the references therein. The world averages based on these analysis can be seen from ref. \cite{hflav}:
\begin{equation}
R(D) = 0.299 \pm 0.003 ,\ \ R(D^*) = 0.258 \pm 0.005, \ \ |V_{cb}|^{exl} = (41.9 \pm 2.0)\times 10^{-3}. 
\label{2017p}
\end{equation}

Uncertainties in $R(D^*)$ are estimated by parametrizing the missing higher-order pieces in the relations between the HQET form-factors \cite{Bigi:2017jbd, Jaiswal:2017rve}. In all of those analyses, the ratios of the HQET form-factors are considered at order ${\cal O}(\alpha_s,\frac{1}{m_b},\frac{1}{m_c})$. In the ref. \cite{Jaiswal:2017rve}, an additional correction of $\approx$ 20\% in the ratios of the HQET form-factors is considered. Also, several different normalizations for the ratios of the form-factors $F_2(w)/F_{i}(w)$ were used to predict $R(D^*)$ and the variations in $R(D^*)$ due to those were noted. Finding a method of predicting $R(D^*)$ which will be relatively less sensitive to the inputs from HQET is essential, and lattice inputs on the form-factors at non-zero recoil are required for that. 

Very recently, Belle has updated their measurement on the decay distributions in $B \to D^*\ell\nu_{\ell}$ \cite{Abdesselam:2018nnh}. They have also extracted the values of $|V_{cb}|$ using the CLN and BGL parametrizations of the form-factors and the results are consistent with each other within the error-bars. The extracted value is lower than what was observed with their 2017 data (eq. \ref{2017p}). These new results from Belle are incorporated in a couple of other analyses \cite{Gambino:2019sif, Bordone:2019vic, Bordone:2019guc} where the authors have updated the prediction of $R(D^*)$ in the SM. 
In both of these analyses, the predicted values of $R(D^*)$ are consistent with the one given in eq. \ref{2017p}, but the central values are lowered by $\approx$ 2\%. In ref. \cite{Bordone:2019vic}, the available results on $B\to D^*$ form-factors from light-cone sum rule (LCSR) \cite{Gubernari:2018wyi} at $q^2 \le 0$ are used. 

In the present article, we have updated our earlier analysis \cite{Jaiswal:2017rve} with the new inputs and have modified the method of our data-analysis. We know that the statistical analyses play an important role in addressing research questions. The two prevalent philosophies in inferential statistics are frequentist and Bayesian. The differences between these two frameworks originate from the way the concept of probability itself is interpreted. In our earlier publication, we had used frequentist framework in analyzing the data, where a parameter of interest is assumed to be unknown, but fixed (has a true value). In general, it is assumed that there is only one true regression coefficient in the population. Here, we have updated our method to the Bayesian view of subjective probability, where all unknown parameters are treated as uncertain and thus should be described in terms of their underlying probability distribution. 

We would like to point out that in our earlier analysis we truncated the BGL series of all the relevant form-factors at $N=2$ which gave us stable results on the extracted $|V_{cb}|$. In such cases, the number of BGL parameters, associated with the three form-factors in $B\to D^*$ decays, will be 9. In a recent work \cite{Bernlochner:2019ldg}, it is shown that at the present level of precision, the optimal number of BGL parameters required to fit the current data is less then 9. However, in such scenarios, the extracted values of $|V_{cb}|$ will be different than the one obtained in the fit with 9 parameters. Therefore, it is clear that truncating the series at $N < 2$ will not be sufficient for the determination of $|V_{cb}|$. Although this is not our main focus, here we have pointed out the use of Akaike Information Theoretic approach (AIC) to find out optimal number of BGL coefficients which can best explain the data. The uses of AIC in the context of NP model selection can be seen from \cite{Bhattacharya:2016zcw,Bhattacharya:2018kig,Bhattacharya:2019dot}
and the references therein. 

Along with the extraction of $|V_{cb}|$ and the prediction of $R(D^*)$, we have extracted a few angular observables related to \bdsttaunu decays. Also, the variation of the form-factors, and the decay rate distributions with the recoil angle $w$ are shown. The results are compared with those obtained from the old Belle data \cite{Abdesselam:2017kjf} and we have noted a shift in the distributions. As before, the updated SM prediction for $R(D^*)$ show deviations from its measured value \cite{hflav}. The presence of new interactions beyond the SM can explain this deviation. Regarding this, there are plenty of analyses available in the literature. Here, we would like to point out a few references where bounds on NP WCs are obtained using the updated results in 2017 \cite{Bhattacharya:2019dot,Feruglio:2018fxo,Huang:2018nnq,Blanke:2018yud,Murgui:2019czp,Asadi:2019xrc,Shi:2019gxi}. In this article, we also extract the new physics Wilson coefficients (WC) using these newly available inputs.

Recently, a set of preliminary results of the HQET form-factors for the $B\to D^* \ell\nu_{\ell}$ decays at non-zero recoil have been presented by Fermilab MILC collaboration \cite{Aviles-Casco:2019zop} and JLQCD \cite{Kaneko:2019vkx}. The analyses have been done with $N_f = 2 + 1$ flavors of sea quarks with variable lattice-spacing. They have also done a chiral-continuum-fit to the available lattice points. Though the error budgets are given, it is not complete. To note the impact of these inputs on the SM prediction of $R(D^*)$, we have done an analysis only with these inputs (without using any experimental data) to extract the form-factors. This type of analysis will give us a clear picture of the experimental biases present in the prediction of $R(D^*)$, which is important, since we neglect the possibility of new physics effects in $b\to c\ell\nu_{\ell}$ (with $\ell=\mu$ or $e$) decays in general. Also, using these lattice inputs, we can obtain the decay rate distributions in \bdstlnu, which can be compared with the corresponding experimental results. Any discrepancies between the two will lead us to reconsider our understanding.

\section{Analysis with the new data}
\label{sec1}
\subsection{Different fits and their comparison}
\label{subsec1a}
\begin{table}[htbp] 
	\begin{center}
		\def\arraystretch{1.2}
		\begin{tabular}{|cccccccc|}
			\hline
			$f_+(w)$ & Value from & \multicolumn{6}{c|}{Correlation} \\
			\& $f_0(w)$ & HPQCD & & & & & & \\
			\hline
		$f_+(1)$ & $1.178 (46)$ & 1. & 0.994 & 0.975 & 0.507 & 0.515 & 0.522 \\
		$f_+(1.06)$ & $1.105 (42)$ &  & 1. & 0.993 & 0.563 & 0.576 & 0.587 \\
			$f_+(1.12)$ & $1.037 (39)$ &  &  & 1. & 0.617 & 0.634 & 0.649 \\
			$f_0(1)$ & $0.902 (41)$ &  &  &  & 1. & 0.997 & 0.988 \\
			$f_0(1.06)$ & $0.870 (39)$ &  &  &  &  & 1. & 0.997 \\
			$f_0(1.12)$ & $0.840 (37)$ &  &  &  &  &  & 1. \\
			\hline
			& Value from & \multicolumn{6}{c|}{} \\
			& MILC & & & & & & \\
			\hline
		$f_+(1)$ & $1.1994 (95)$ & 1. & 0.967 & 0.881 & 0.829 & 0.853 & 0.803 \\
		$f_+(1.08)$ & $1.0941 (104)$ &  & 1. & 0.952 & 0.824 & 0.899 & 0.886 \\
			$f_+(1.16)$ & $1.0047 (123)$ &  &  & 1. & 0.789 & 0.890 & 0.953 \\
			$f_0(1)$ & $0.9026 (72)$ &  &  &  & 1. & 0.965 & 0.868 \\
			$f_0(1.08)$ & $0.8609 (77)$ &  &  &  &  & 1. & 0.952 \\
			$f_0(1.16)$ & $0.8254 (94)$ &  &  &  &  &  & 1. \\     
			\hline    
		\end{tabular}
	\end{center}
	\caption{Lattice QCD results for $f_+$ and $f_0$ for different values of $w$. The upper half of the table have been obtained using the fit results from the HPQCD collaboration \cite{Na:2015kha}, and the lower half are the results obtained by the Fermilab MILC collaboration \cite{Lattice:2015rga}.}
	\label{tab:latinput}
\end{table}

\begin{table}[htbp]
	\begin{center}
	\def\arraystretch{1.2}
		\begin{tabular}{|c|c|}
			\hline
			$\chi^L_{0^+}(0) = 6.204 \times 10^{-3}$ \\
			$\tilde\chi^L_{0^-}(0) = 19.421 \times 10^{-3} $  \\
			$\tilde\chi^T_{1^-}(0) = 5.131 \times 10^{-4}  GeV^{-2}$   \\
			$\chi^T_{1^+}(0) = 3.894 \times 10^{-4}  GeV^{-2} $  \\
			\hline
		\end{tabular}
	\end{center}
	\caption{Various inputs in our analysis, the $\chi$'s are the functions relevant for BGL parametrizations of the form-factors, for detail see \cite{Boyd:1997kz,Bigi:2017jbd}}
	\label{tab:theoryinputs}
\end{table}

\begin{table}[t]
	\begin{center}
		\def\arraystretch{1.5}
		\begin{tabular}{|c|c|c|c|c|c|}
			\hline
			&No. of estimated& &&& \\
			$(n_f, n_{F_1}, n_g)$ & parameters &p-Value (\%) & $AIC_c$ & $\Delta AIC_c$ & $|V_{cb}|\times 10^{3}$ \\
			& including $|V_{cb}|$ & &&& \\
			\hline
			(1,2,0) & 6 & 56.95 & 47.38 & 0     & 39.76 \errvec{87 \\ 98} \\
			\hline
			(2,2,0) & 7 & 52.16 & 50.29 & 2.91  & 39.79 \errvec{111 \\ 99}  \\
			\hline
			(1,2,1) & 7 & 51.13 & 50.50 & 3.12  & 39.38 \errvec{122 \\ 119}  \\
			\hline
			(1,2,2) & 8 & 46.75 & 53.50 & 6.12  & 39.50 \errvec{118 \\ 107}  \\
			\hline
			(2,2,1) & 8 & 46.19 & 53.61 & 6.23  & 39.35 \errvec{115 \\ 116}  \\
			\hline
			(1,1,0) & 5 & 22.28 & 53.84 & 6.47  & 40.35 \errvec{86 \\ 98} \\
			\hline
			(2,1,0) & 6 & 22.31 & 55.50 & 8.12  & 40.00 \errvec{100 \\ 101} \\
			\hline
			(1,1,1) & 6 & 21.38 & 55.79 & 8.42  & 39.85 \errvec{107 \\ 106} \\
			\hline
			(2,2,2) & 9 & 41.81 & 56.80 & 9.43  & 39.37 \errvec{107 \\ 121} \\
			\hline
		\end{tabular}
	\end{center}
	\caption{Ranking of various $(n_f, n_{F_1}, n_g)$ scenarios with $\Delta AIC_c \lsim 10$. For details, see the text.}
	\label{tab:modelselection}
\end{table}

\begin{table}[htbp]
	\begin{center}
	\def\arraystretch{1.5}
   \begin{adjustbox}{width=\textwidth}
	\begin{tabular}{|c|c|c|c|}
		\hline
		&$n_f=1, n_{F_1}=2, n_g=0$&\multicolumn{2}{|c|}{$n_f = n_{F_1} = n_g = N = 2$ } \\
		\cline{2-4}
		Parameters&  Belle 2019 data + &  Belle 2019 data + &  Belle 2019 data +  LCSR at $q^2=0$ \cite{Gubernari:2018wyi} \\
		&$h_{A_1}(1)$ from MILC \cite{Bailey:2014tva}&$h_{A_1}(1)$ from MILC \cite{Bailey:2014tva}&  + $h_{A_1}(1)$ from MILC \cite{Bailey:2014tva}\\
		\hline
		$|V_{cb}|\times 10^{3}$ & 39.76 \errvec{87 \\ 98}  &39.37 \errvec{107 \\ 121} & 39.56 \errvec{104 \\ 106}\\
		\hline
		$a^f_0$& 0.0122 (2)  & 0.0122 (2) & 0.0122 (2)\\
		$a^f_1$  & 0.0056 \errvec{54 \\ 58}  & 0.0012 \errvec{171 \\ 286} & 0.0256 \errvec{232 \\ 168}\\
		$a^f_2$   & $-$ & 0.0792 (5474) & -0.3108 (4175)\\
		\hline
		$a^{F_1}_1$& 0.0062 \errvec{20 \\ 21}  & 0.0068 \errvec{22 \\ 24}  & 0.0063 \errvec{21 \\ 19}\\
		$a^{F_1}_2$ &-0.1128 \errvec{333 \\ 352} & -0.1157 \errvec{412 \\ 392} & -0.1033 \errvec{323 \\ 366}\\
		\hline
		$a^g_0$ & 0.0268 \errvec{7 \\ 8}  & 0.0289 \errvec{81 \\ 126} & 0.0272 \errvec{42 \\ 47}\\
		$a^g_1$  & $-$ & -0.0840 \errvec{1812 \\ 1343} & -0.0088 \errvec{1199 \\ 1031}\\
		$a^g_2$ & $-$ & -0.0016 (5304) & 0.0013 (5699) \\
		\hline
	\end{tabular}
\end{adjustbox}
	\end{center}
	\caption{The results of the fit to new Belle data \cite{Abdesselam:2018nnh} in \bdstlnu decay using BGL parametrization of the form-factors for $n_f=1, n_{F_1}=2, n_g=0$ (second column) and for $N=2$ (third column). The last (fourth) column corresponds to the fit to new Belle data \cite{Abdesselam:2018nnh} and LCSR inputs (for $q^2=0)$ \cite{Gubernari:2018wyi} for $N=2$. } 
	\label{tab:BGLfit}
\end{table}

Essentially, there are two form-factors $f_+(z)$ and $f_0(z)$ relevant for \bdlnu decays, while those for \bdstlnu decays are given by $f(z)$, $g(z)$, $F_1(z)$ and $F_2(z)$, respectively. Following the BGL parametrization, each of these form factors can be written as a series expansion in $z$,
 \begin{equation}
	F_i(z) = \frac{1}{P_{i}(z )\phi_{i}(z )}\sum_{n=0}^N a_n^{F_i}\ z^n,
	\label{bglform}
\end{equation}
 with
\begin{equation}
	z=\frac{\sqrt{w +1}-\sqrt{2}}{\sqrt{w +1}+\sqrt{2}}, 
\end{equation}
where $w$ is the recoil angle. The mathematical forms of $\phi_i$'s and the Blaschke factor $P_i(z)$ can be seen from \cite{Boyd:1997kz}. The numerical values of the relevant $\chi$ functions, associated with the form-factors, are given in table \ref{tab:theoryinputs}, with details given in ref. \cite{Boyd:1997kz,Bigi:2017jbd,Jaiswal:2017rve}. Here, $F_i(z)$ include all the relevant form-factors $f_+(z)$, $f_0(z)$, $F_1(z)$, $f(z)$, $g(z)$ and $F_2(z)$ respectively. For the coefficients $a_n^{F_i}$, we are using the weak unitarity constraints. Here, $z$ is a kinematic variable and for the semileptonic decays under consideration, its values lie between 0 and 0.0456; for details, see \cite{Boyd:1997kz}. The form factors in  $B \to D^{(*)} \ell \nu$ decays can be fully expressed in terms of heavy quark effective theory (HQET) form-factors $h_+(w), h_-(w), h_V(w), h_{A_1}(w), h_{A_2}(w)$ and $h_{A_3}(w)$. We have simultaneously extracted $|V_{cb}|$ and the form-factors $f(z)$, $F_1(z)$, and $g(z)$ from the fit\footnote{All numerical analysis in this work has been done using Optex \cite{optex}, a Mathematica package.} to the available new data on the differential rates and angular distributions in $B\to D^{*}\ell \nu_{\ell}$ \cite{Abdesselam:2018nnh} (and using the lattice input on the form factor $h_{A_1}(1) = 0.906(13)$ from the unquenched Fermilab/MILC lattice data \cite{Bailey:2014tva}). Note that the Belle 2019 data does not provide enough information required for a full unfolding of the data. However, they have provided all the necessary information to perform a fit to the form-factors. For details, see section VIII of ref. \cite{Abdesselam:2018nnh}. The $\chi^2$ function is defined as 
	\begin{equation}
	\chi^2 = \sum_{i,j} \big(N_i^{obs} - N_i^{exp}\big) C_{ij}^{-1} \big(N_j^{obs} - N_j^{exp}\big), 
	\end{equation}
	where $C_{ij}^{-1}$ is the inverse of the covariance matrix, and $N_i^{obs}$ and $N_i^{exp}$ are the observed and expected number of events in the $i^{th}$ bin
	with $N_i^{exp} = \sum_{j=1}^{40} R_{ij} \epsilon_j N_j^{theory}$. Here, $R$ is the response matrix, and $N_j^{theory}$ can be obtained from the related theory expressions. $R$ can be found in the reference mentioned above. The background-subtracted signal yield $N^{obs}$ with the statistical error and the corresponding signal efficiencies ($\epsilon$) for all the 40 bins are there too, along with all other necessary information, e.g., systematic uncertainty, statistical and systematic correlation matrices. Thus, instead of comparing the actual theory with the unfolded data, we are comparing a predicted theory at the level of the smearing/folding with the actual observed data. This is because folding an assumed true distribution is simpler than unfolding an observed distribution in an attempt to obtain the true one, and hence, the preferred procedure in general.
We have made the necessary corrections to avoid the bias due to the D'Agostini effect \cite{DAgostini:1993arp} while using the experimental systematic uncertainties in our analysis. Some earlier analyses\cite{Gambino:2019sif, Jung:2018lfu} on $B\to D^{(*)} l \nu$ have also incorporated these corrections. Using these fit results and the inputs given in table \ref{tab:latinput}, we have predicted $R(D^{*})$, and other relevant observables like the tau-polarization $P_{\tau}(D^*)$, $D^*$-polarization $F_L(D^*)$, and the forward-backward asymmetry $A_{FB}(D^*)$. Since there are no new updates on $B\to D\ell \nu_{\ell}$ decay, we do not repeat the analysis of this decay mode.

\begin{figure}[t]
	\begin{center}
	\includegraphics[width=0.95\linewidth]{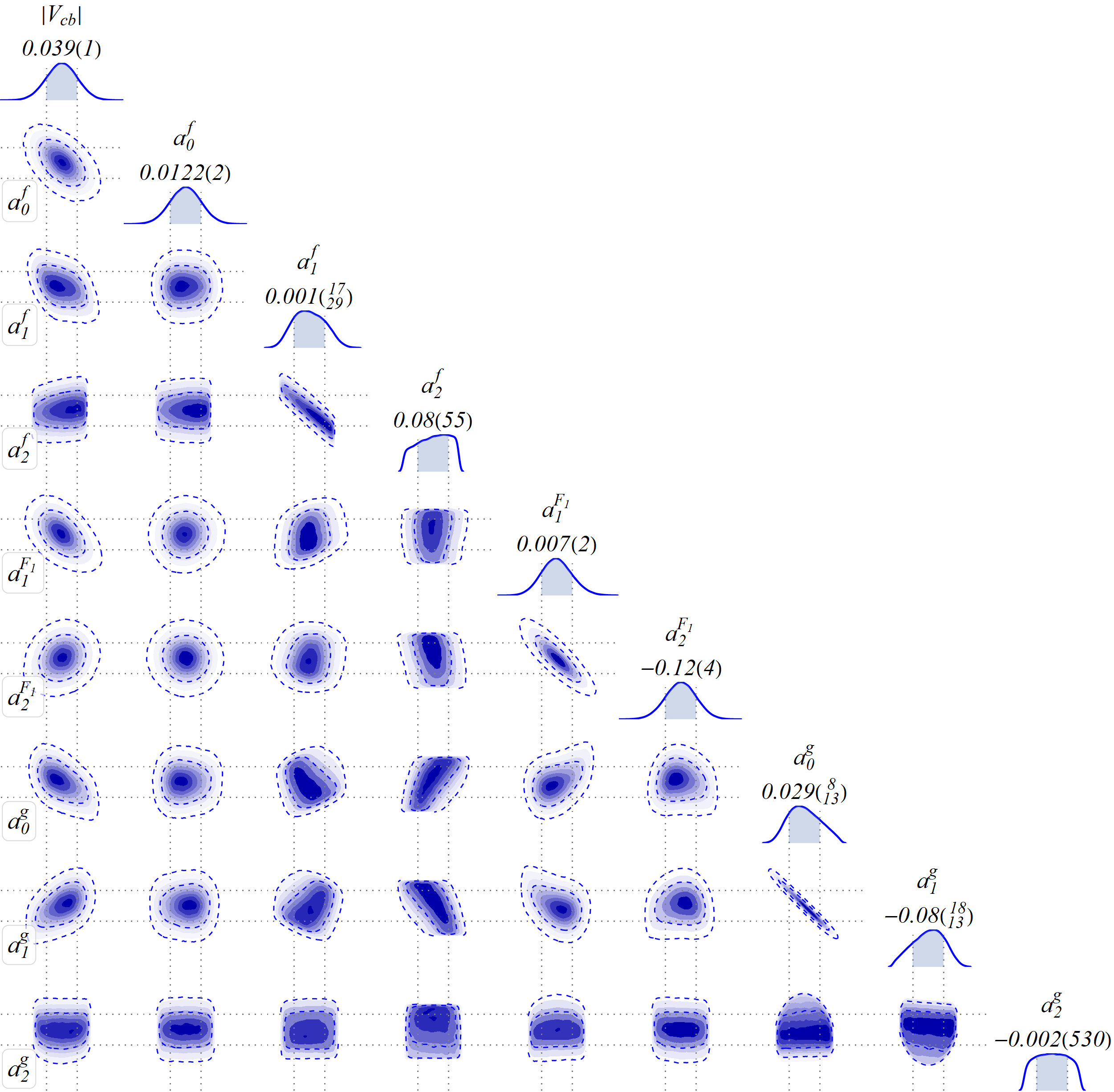}
	\end{center}
		\caption{The Bayesian fit results for all the BGL parameters of \bdstlnu form-factors corresponding to $N=2$ obtained with 2019 Belle data (third column of table \ref{tab:BGLfit}). The plot shows 1D and 2D (depicting the correlations) marginal probability distributions and the best-fit values for the parameters.}
	\label{fig:parcorplot}
\end{figure}%

We have analyzed the data using different orders of truncation in the series expansion of the form-factors and compared them. Though our main focus is on the BGL form-factors truncated at $N=2$, i.e., $(n_f, n_{F_1}, n_g) \equiv (2,2,2)$ \footnote{$n_{\scaleto{F_i}{5pt}}$ denotes the order at which the BGL expansion for the form factor `$F_i$' is truncated.}, we perform several other Bayesian fits by varying the order of truncation for each of the BGL form factors from 0 to 2 independently to study their sensitivity to the present data. We have considered all possible combinations of the values of $n_f, n_{F_1}$ and $n_g$ ranging between 0 and 2. Since the Fermilab/MILC lattice input on $h_{A_1}(1)$ \cite{Bailey:2014tva} precisely constraints the BGL coefficient $a^f_0$ in the fit and the coefficient $a^{F_1}_1$ is eliminated from the fit using the relation $F_1(w=1)=(m_B-m_{D^*})f(w=1)$, the minimal scenario considered here is $(n_f, n_{F_1}, n_g) \equiv (1,1,0)$. With the maximal scenario being $(n_f, n_{F_1}, n_g) \equiv (2,2,2)$, we get a total of 12 possible combinations. We then use the second-order variant of Akaike's Information Criterion $(AIC_c)$ to do a data-based comparison and ranking of all the scenarios in hand. We use the mean values for the parameters from their 1D marginal probability distributions to calculate the $(AIC_c)$. The ranking is based on the values of $AIC_c$ or $\Delta AIC_c$ and the scenario with the lowest $AIC_c$ will top the list. The outcome of this analysis is presented in table~\ref{tab:modelselection} where we have only shown the scenarios with $\Delta AIC_c \lsim 10$. We find that $(n_f, n_{F_1}, n_g) \equiv (1,2,0)$ is the only scenario with $\Delta AIC_c < 2$ and thus, this combination forms the best scenario to explain the present Belle data. Note that the scenario (2,2,2) is much below in the $AIC_c$ ranking, which could be an indicator that the precision of present data is not sufficient for the precise extraction of the nine parameters in this scenario. 

In all the Bayesian fits involving BGL parameters hereafter, we have implemented the weak unitarity constraints as a posterior requirement, which have been neglected in ref. \cite{Bernlochner:2019ldg}. These unitarity bounds are imposed as a hard cut-off on the posterior distributions by the introduction of appropriate penalty functions using Lagrange multipliers of quadratic nature. The Bayesian fit results for $|V_{cb}|$ and \bdstlnu form-factor parameters for the scenarios $(1, 2, 0)$ and $(2, 2, 2)$ are given in the second and third columns of table \ref{tab:BGLfit} respectively. As expected, the extracted parameters in the best possible scenario have smaller uncertainties compared to those in the scenario $(2,2,2)$. Also, the coefficients/parameters of the expansion which are dropped in $(1,2,0)$-scenario have large uncertainties in the scenario $(2,2,2)$, which points to the fact that the present data is not sensitive enough to extract the higher powers of the expansion. The marginal probability distributions for the BGL coefficients of all the \bdstlnu form-factors corresponding to the $N=2$ Bayesian fit and the correlations between them are depicted in Figure \ref{fig:parcorplot}. We note that the 1D marginal distribution of the second-order coefficients of the BGL expansion are almost flat: indicative of the poor sensitivities of these coefficients toward the present data.  
  
Very recently, there are updates from LCSR on the form-factors $V(q^2), A_1(q^2), A_2(q^2)$ and $A_0(q^2)$ at $q^2 = 0$ \cite{Gubernari:2018wyi}. These QCD form-factors can be expressed in terms of the BGL form-factors, $g, f, F_1$ and $F_2$. We have utilized these newly available LCSR results along with the other inputs and performed a Bayesian fit. The results are presented in the fourth column of table \ref{tab:BGLfit}. The LCSR inputs are also available for a few other values of $q^2 = -5$, $-10$, $-15$ (in ${\it {GeV}^2}$). However, we have not used these inputs in our analysis because of the reason stated in the following. The BGL parametrization of the form-factors rely on a Taylor series expansion about $z=0$. The key ingredient in this approach is the transformation that maps the complex $q^2$ plane onto the unit disc $|z| \le 1$. Therefore, small values of $z$ ensure faster convergence of the series. Now, for the semileptonic decay $B \to D^{\ast} l \nu$, the kinematically allowed region is $0 < z \le 0.0456$. In this semileptonic region, the maximum value of $z$ is obtained at $q^2=0 (GeV^2)$, and thus the large negative values of $q^2$ lead to relatively larger values of $z$ for which the higher-order terms in the BGL expansion may become important. We simply tried to avoid such consequences.  Hence, we mostly concentrate on the LCSR inputs for $q^2 = 0$ for the BGL form-factor fits. The extracted $|V_{cb}|$ is consistent with the one obtained without LCSR inputs.

\begin{figure}[t]
	\begin{center}
		\subfigure[]{%
			\label{fig:fplot}
			\includegraphics[width=0.45\textwidth]{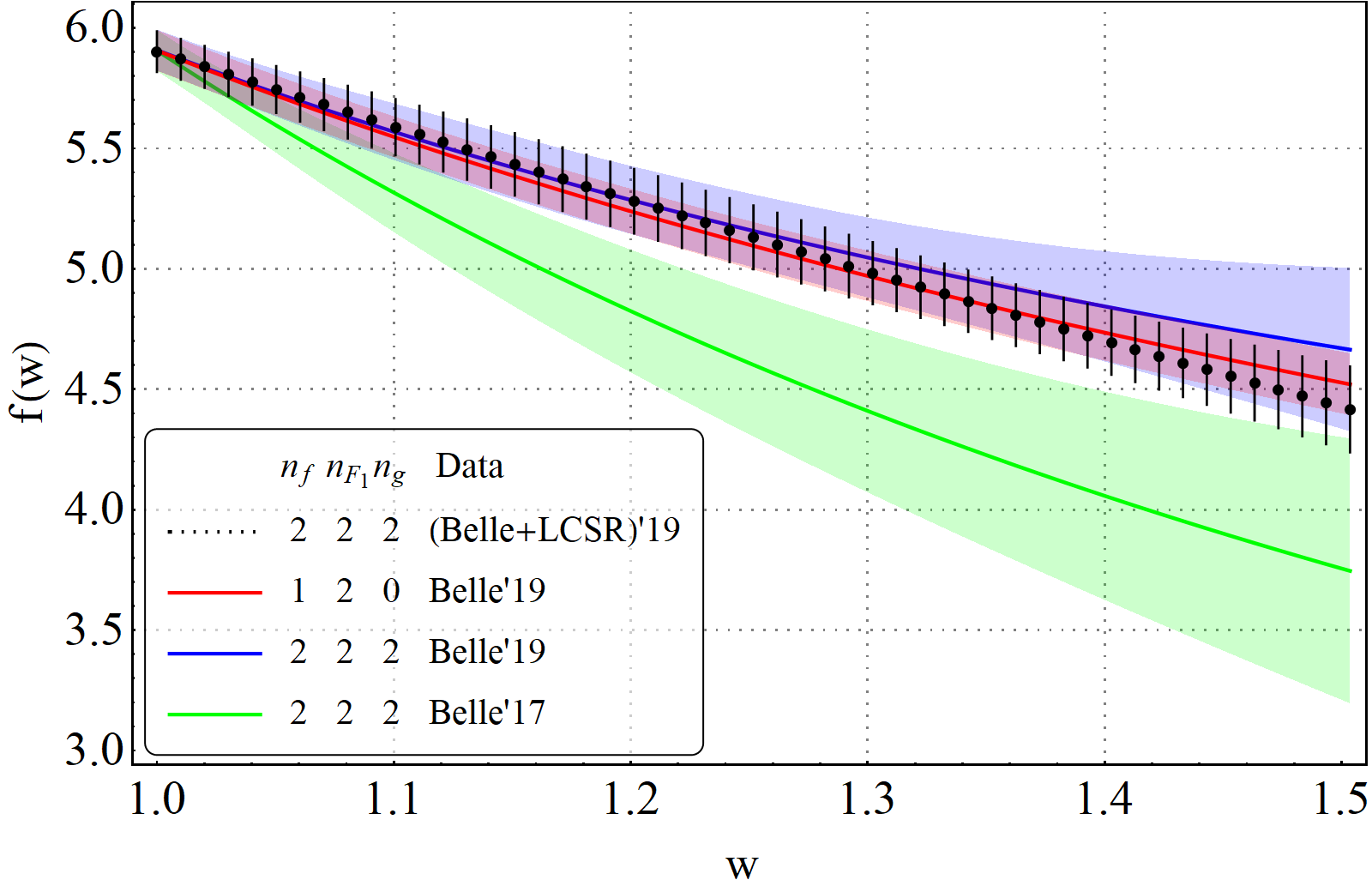}
		}%
	    ~~~~
       \subfigure[]{%
			\label{fig:f1plot}
			\includegraphics[width=0.45\textwidth]{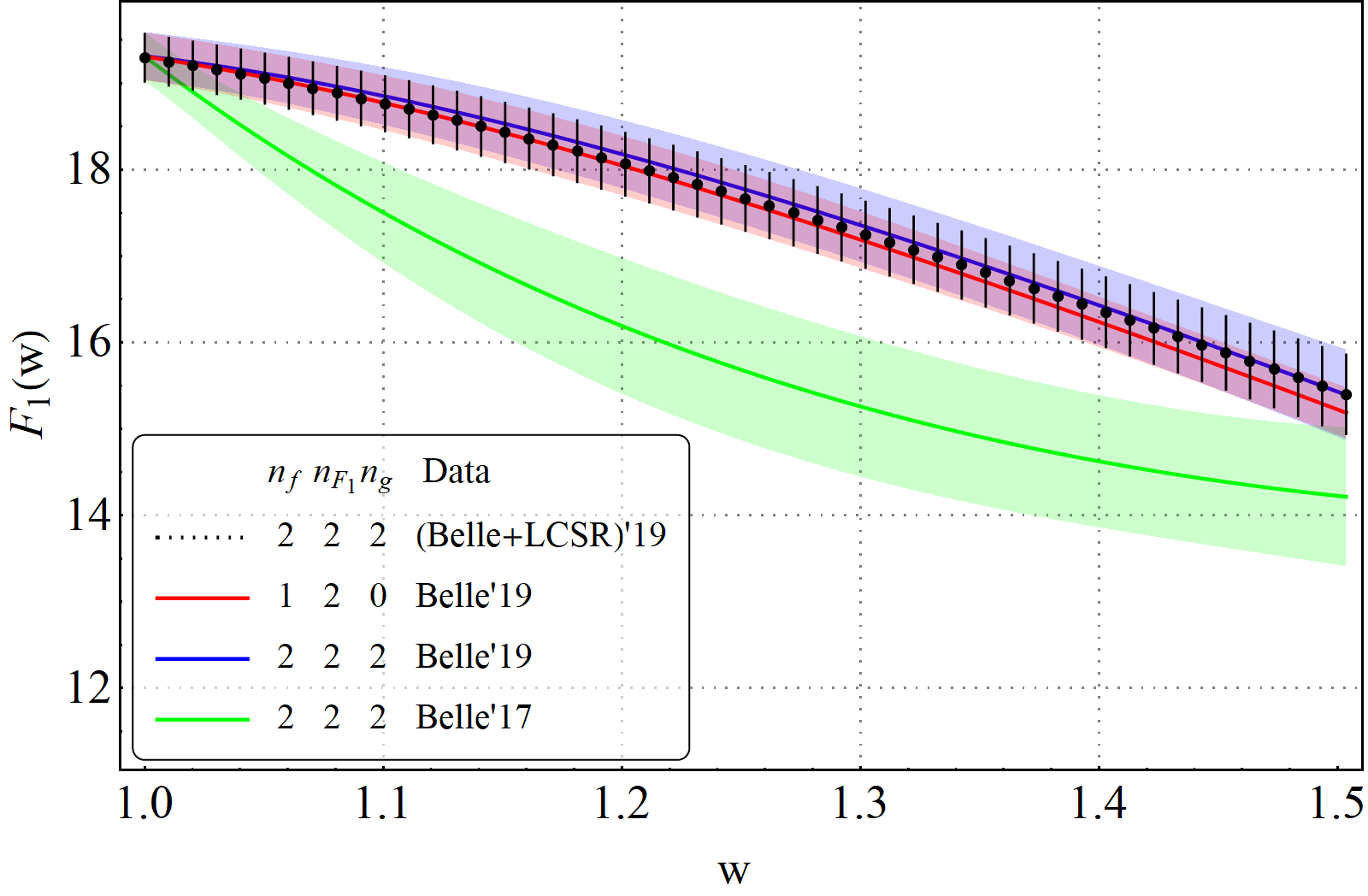}
		}%
		\\
		\subfigure[]{%
			\label{fig:gplot}
			\includegraphics[width=0.45\textwidth]{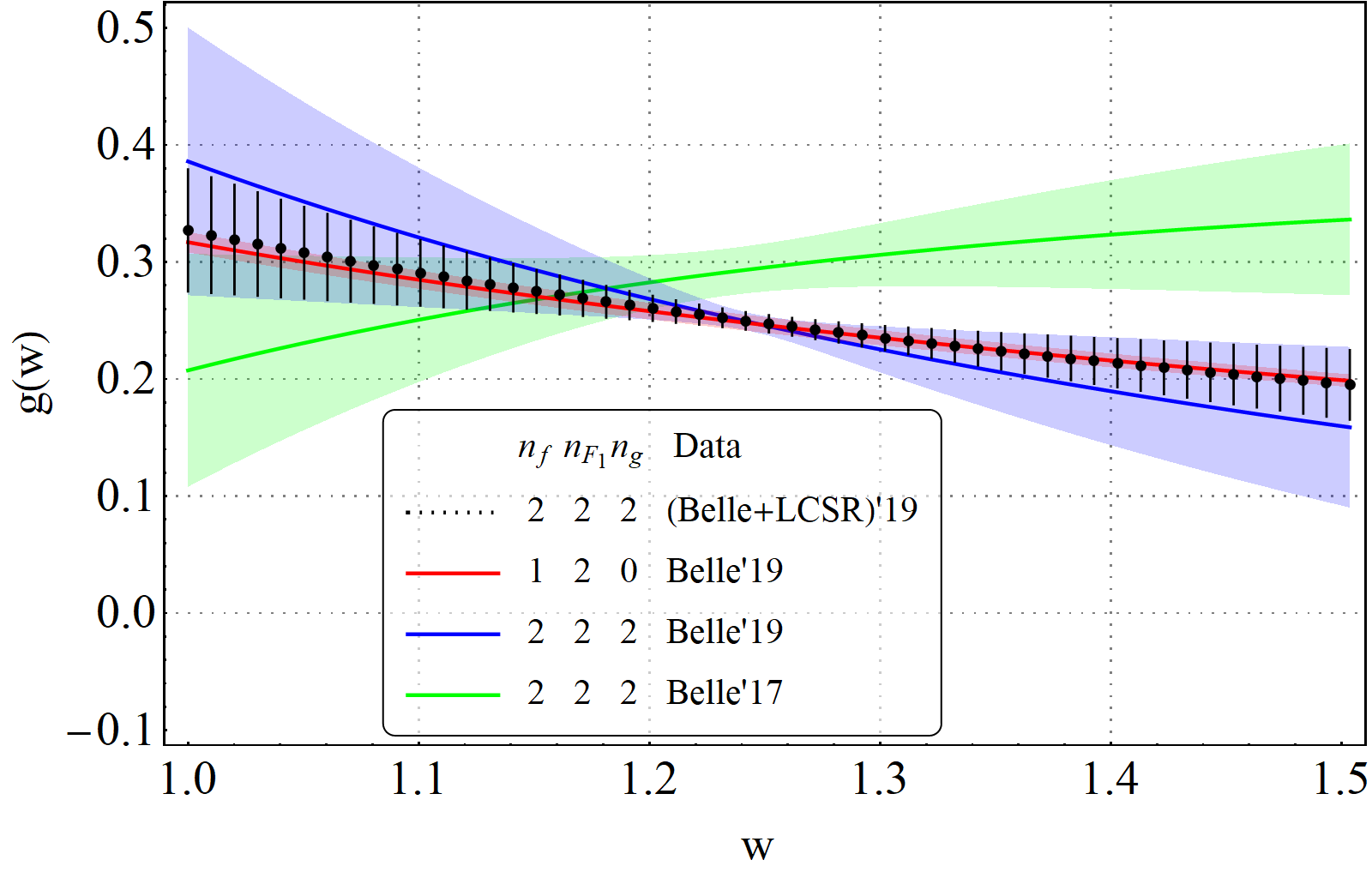}
		}%
	\end{center}
\caption{The shape of \bdstlnu form-factors obtained from the $N=2$ fit to new Belle data and LCSR inputs (black bars) is compared to those obtained with new Belle data for $N=2$ (blue band) and for $n_f=1, n_{F_1}=2, n_g=0$ (red band), and also to the one obtained using old Belle data (green band) for $N=2$.}
\label{fig:ffplots}
\end{figure}

The $w$-distributions of the form-factors in \bdstlnu, extracted from different fits in table \ref{tab:BGLfit}, are shown in figure \ref{fig:ffplots}.
It compares the $w$ dependence of the three \bdstlnu form factors, $f(w), F_1(w)$ and $g(w)$ obtained from three different Bayesian fits.
The effects of the choices of experimental and theory inputs and the order of truncation of BGL expansion can be seen in these figures. There are differences between the results obtained using the 2017 \cite{Abdesselam:2017kjf} and the 2019 Belle results \cite{Abdesselam:2018nnh}, for all the form-factors. These differences become more prominent in the large-recoil limit, since the lattice results play an important role in constraining the BGL coefficients at zero-recoil. Also, the $w$-dependence of the respective form-factors are consistent with each other for both $(1,2,0)$ and $(2,2,2)$ scenarios. As expected, the errors are a little less in the scenario $(1,2,0)$. The $w$-dependence of all the form-factors are unaltered after incorporating the inputs from LCSR at maximum recoil while the respective uncertainties are reduced.         

\subsection{$R(D^*)$ in the SM}
\label{subsec1b}

\begin{table}[!htbp]
	\begin{center}
		\def\arraystretch{1.2}
				\begin{adjustbox}{width=\textwidth}
		\begin{tabular}{|c|c|c|c|c|c|c|}
			\hline
			Parameters &  $\eta(1)$ & $\eta^{\prime}(1)$ & $\chi_2(1)$ & $\chi^{\prime}_2(1)$ &$\chi^{\prime}_3(1)$ & $\Delta_{\mp}$ \\
			\hline
			&&&&&&\\
			Values & 0.373 \errvec{89 \\ 58} & -0.060 \errvec{41 \\ 33} & -0.059 (20) & -0.003 \errvec{19 \\ 21} & 0.037 \errvec{18 \\ 19} & 0.91 (21) \\
			&&&&&&\\
			\hline
		\end{tabular}
	\end{adjustbox}
	\end{center}
	\caption{The fit results for the sub-leading Isgur-Wise functions and $\Delta_{\mp}$}
	\label{tab:hqetfit}
\end{table}

We also obtain estimates for $R(D^*)$ and other $B \to D^* \tau \nu_{\tau}$ observables for all the fits; as listed in table \ref{tab:BGLfit} for $N=2$. As mentioned earlier, there will be one additional form-factor $F_2(z)$ in \bdsttaunu, which we can not constrain from the experimental data in \bdstlnu. We estimate the parameters of $F_2(z)$ by exploiting the HQET relations between the form-factors as shown in our earlier work \cite{Jaiswal:2017rve}. In the following, we will briefly describe the method; for more details, see \cite{Jaiswal:2017rve}.  


The ratio $f_+(w)/f_0(w)$ can be expressed in HQET up to order $\mathcal{O}(\alpha_S)$ in perturbative corrections and $\mathcal{O}(\Lambda_{QCD}/m_{b,c})$ in non-perturbative corrections \cite{Caprini:1997mu,Bernlochner:2017jka}, which are expressed in terms of a few sub-leading Isgur-Wise functions:  $\eta(1),~\eta^{\prime}(1),~\chi_2(1),~\chi^{\prime}_2(1)$, and $\chi^{\prime}_3(1)$. We use the lattice inputs on \bdlnu (table \ref{tab:latinput}) to create synthetic data-points for the HQET fit to the sub-leading Isgur-Wise functions. For example, we consider the ratio $f_+(w)/f_0(w)$ for $w = 1, 1.08, 1.16$, using lattice data from MILC and for $w = 1, 1.03, 1.06, 1.09, 1.12$, we use lattice data from HPQCD. Other relevant inputs like the quark masses ($m_b, m_c$), $\alpha_S$ and $\Lambda_{QCD}$ are taken from the reference \cite{Bernlochner:2017jka}. We use these eight synthetic data-points to perform a Bayesian fit to the sub-leading Isgur-Wise functions. We observe that $\chi_2(1),~\chi^{\prime}_2(1)$, and $\chi^{\prime}_3(1)$ are relatively insensitive to the form factor ratios used in the fit and hence, we use the QCDSR predictions \cite{Neubert:1992wq,Neubert:1992pn} for $\chi_2(1),~\chi^{\prime}_2(1)$, and $\chi^{\prime}_3(1)$ to specify the prior distributions for these parameters in the Bayesian fit.  

In addition, to account for the missing higher order corrections, several normalizing parameters $(\Delta$s$)$ have been introduced here in the ratios of the HQET form factors. We just need to make the following replacements in order to be able to estimate the size of these higher order corrections:
\begin{align}
\frac{h_{v}}{h_{A_1}} &\to \frac{h_{v}}{h_{A_1}} \Delta_v, \ \   \frac{h_{A_3}}{h_{A_1}} \to \frac{h_{A_3}}{h_{A_1}} \Delta_{31}, \ \  
\frac{h_{A_2}}{h_{A_1}} \to \frac{h_{A_2}}{h_{A_1}} \Delta_{21},\ \   \frac{h_{-}}{h_{+}} \to \frac{h_{-}}{h_{+}} \Delta_{\mp}, \ \   \frac{h_{+}}{h_{A_1}} \to \frac{h_{+}}{h_{A_1}} \Delta\,.
\end{align}
For the ratio $f_+(w)/f_0(w)$, the only normalizing parameter involved is $\Delta_{\mp}$. We use the conservative estimate for $\Delta_{\mp} = 1 \pm 0.2$ to define its prior distribution in the Bayesian fit. The fit results for the sub-leading Isgur-Wise functions and $\Delta_{\mp}$ are summarized in table \ref{tab:hqetfit}.

\begin{table}[htbp]
	\begin{center}
		\def\arraystretch{1.5}
			\begin{tabular}{|c|c|c|cccc|}
				\hline
				Inputs used in the fit & Observable & Value & \multicolumn{4}{c|}{Correlations}  \\
				\hline
				  & $R({D^*})$        & 0.256 \errvec{7 \\ 8}  &1&0.639&0.448&0.354  \\   
				 Belle 2017 \cite{Abdesselam:2017kjf} +& $P_{\tau}^{({D^*})}$& -0.485 \errvec{26 \\ 28} &&1&0.596&0.554   \\
				 $h_{A_1}(1)$ from MILC \cite{Bailey:2014tva}& $F_L^{D^*}$& 0.458 \errvec{15 \\ 17} &&&1&0.742 \\
				 &$A_{FB}^{({D^*})}$ & -0.033 (24)	&&&&1 \\
				\hline
				  & $R({D^*})$        & 0.251 \errvec{4 \\ 5} &1&0.944&0.839&0.617 \\
				 Belle 2019 \cite{Abdesselam:2018nnh} +& $P_{\tau}^{({D^*})}$& -0.492 \errvec{25 \\ 24} &&1&0.830&0.572 \\
				 $h_{A_1}(1)$ from MILC \cite{Bailey:2014tva}& $F_L^{D^*}$& 0.469 \errvec{10 \\ 11} &&&1&0.764\\
				 &$A_{FB}^{({D^*})}$ & -0.038 \errvec{22 \\ 21} 	&&&&1 \\
				\hline
				Belle 2019 \cite{Abdesselam:2018nnh} + & $R({D^*})$ & 0.252 \errvec{6 \\ 7} &1&0.978&0.942&0.886 \\
				 LCSR at $q^2=0$\cite{Gubernari:2018wyi} +& $P_{\tau}^{({D^*})}$& -0.490 \errvec{32 \\ 36} &&1&0.944&0.886 \\
				 $h_{A_1}(1)$ from MILC \cite{Bailey:2014tva}& $F_L^{D^*}$& 0.469 \errvec{14 \\ 15} &&&1&0.921 \\
				 &$A_{FB}^{({D^*})}$ & -0.026 (20)	&&&&1 \\
		\hline
			\end{tabular}
	\end{center}
	\caption{SM predictions for and the correlations between the observables in $B\to D^{*}\tau \nu_{\tau}$ decays as obtained from different fit scenarios for $N=2$. The estimates corresponding to the 2017 Belle data correspond to the analysis in our earlier work. \cite{Jaiswal:2017rve}}
	\label{tab:SMpred}
\end{table}

\begin{figure}[!htbp]
	\begin{center}
		\includegraphics[width=0.65\linewidth]{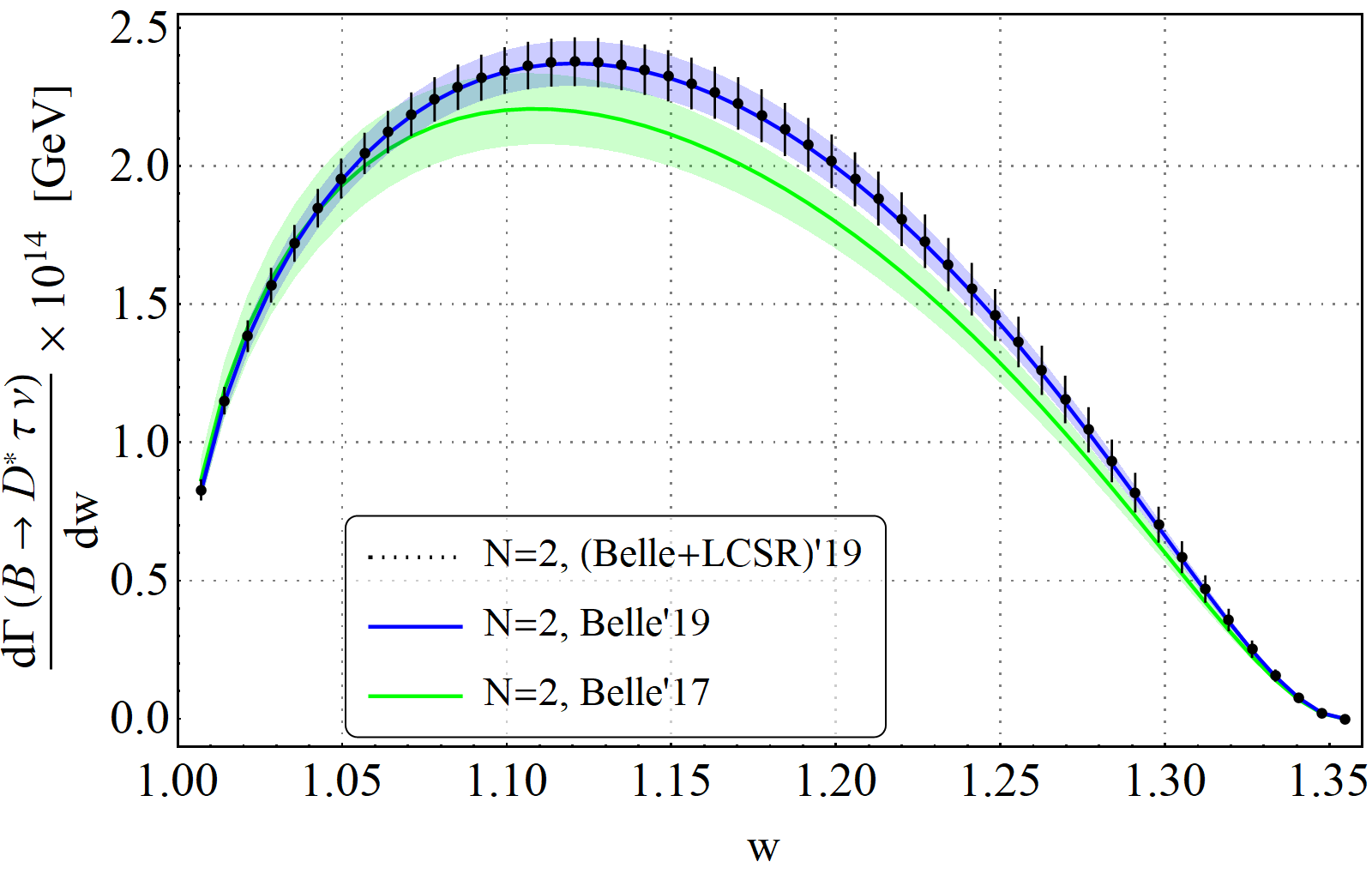}
	\end{center}
	\caption{Comparison between \bdsttaunu decay distributions obtained from different fit scenarios discussed in table \ref{tab:SMpred}. This plot uses the same colour-coding as the form-factor plots in figure \ref{fig:ffplots}. }
	\label{fig:ddplot}
\end{figure}%

Finally, using the following equation:
\begin{equation}
 \left(\frac{F_2(w)}{f_{+,0}(w)}\right)_{\scaleto{BGL}{5pt}} =  \left(\frac{F_2(w)}{f_{+,0}(w)}\right)_{\scaleto{HQET}{5pt}}\,, 
\label{synthetic}
\end{equation}
we express $F_2(w)/f_{+,0}(w)$ in terms of the sub-leading Isgur-Wise functions. We have created synthetic data points for the right hand side of the above equation using the fit results given in table \ref{tab:hqetfit}. Three more normalizing parameters, $\Delta_{21}, \Delta_{31}$ and $\Delta$, come into play here in defining these ratios and we use their conservative estimates: $\Delta_{21} = 1 \pm 0.2$, $\Delta_{31} = 1 \pm 0.2$, and $\Delta = 1 \pm 0.1$. As the BGL coefficients for the form factors $f_+$ and $f_0$ are already known from the \bdlnu fit \cite{Jaiswal:2017rve}, the only unknowns are the BGL coefficients of $F_2$ for $N=2$, i.e. $a^{F_2}_0$, $a^{F_2}_1$, and $a^{F_2}_2$. We use the QCD relation $A_0(q^2=0) = A_3(q^2=0)$, which is equivalent to
\begin{equation}
F_2(q^2=0) = \frac{2 F_1(q^2 = 0)}{m^2_B - m^2_{D^*}}\,,
\label{qcdrelation}
\end{equation}
to eliminate $a^{F_2}_1$ (for details, see \cite{Jaiswal:2017rve}). Thus, in an attempt to obtain a conservative estimate of the form factor $F_2$, we fit the BGL parameters $a^{F_2}_0$ and $a^{F_2}_2$ to the above-generated synthetic data-points. 

As we find that this Bayesian fit is highly insensitive to the BGL coefficient $a_2^{F_2}$, we have set $a_2^{F_2}=0$ without any loss of generality in making the predictions of the  $B\to D^* \tau \nu_{\tau}$ observables. Thus using the equation \ref{synthetic} for $f_+$ with $w=1$, we obatin $a^{F_2}_0=0.056(8)$. Since, BGL \bdlnu fit (for details, see ref. \cite{Jaiswal:2017rve}) is highly constrained by the lattice data from MILC we can safely use the MILC data in estimating $a^{F_2}_0$. 
Using this conservative estimate of $a^{F_2}_0$ together with the fit results for the BGL \bdstlnu form-factor coefficients for $N=2$, we have estimated $R({D^*})$, $P_{\tau}(D^*)$, $F_L(D^*)$, and $A_{FB}(D^*)$, which are tabulated in table \ref{tab:SMpred}. The predictions for all the observables (68\% credible intervals around the mode\footnote{The mode is the value which occurs most frequently in a data set. In case of skewed distributions, the mode is different from the mean; however, for Gaussian distribution, mode and mean are the same.}) and their correlations, here and thereafter, are obtained using samples of the parameteric posterior distributions from the Bayesian fits. All these predictions are consistent with the previously obtained results given in \cite{Bhattacharya:2018kig,Gambino:2019sif,Bordone:2019guc,Bordone:2019vic}. Note that the prediction for $P_{\tau}(D^*)$ is consistent with measurement \cite{Hirose:2016wfn}. On the other hand, $F_L^{D^*}$ is only consistent with the corresponding measurement \cite{Abdesselam:2019wbt} at 2$\sigma$. Also, the SM prediction is lower than the experimental result. For the detailed mathematical expressions of these angular observables, see \cite{Sakaki:2013bfa}. The estimates are given only for the analysis with the BGL coefficients at order $N=2$. The second and third rows of the table \ref{tab:SMpred} show the predictions for the same observables using Belle 2017 and Belle 2019 data, respectively. We note that all the predicted values are consistent with each other within the error-bars. While we note a reduction in uncertainty in all the observables for the present data set, the central value of $R(D^*)$ has reduced by $\approx$ 2\%. 

The fourth/last row of table \ref{tab:SMpred} represents the results which are obtained using the fit results with LCSR inputs at $q^2=0$. Here, we extract $a^{\mathcal{F}_2}_0$ as before, but $a^{\mathcal{F}_2}_1$ is constrained using the LCSR input, $A_0(q^2=0)=  0.68 \pm  0.18$ \footnote{This numerical value has been obtained using the values of $A_1(0)$ and  $A_2(0)$ from \cite{Gubernari:2018wyi} see also \cite{Faller:2008tr}}, not from the equation \ref{qcdrelation}. The modern LCSR technique uses the correlation functions, which are expanded near the light-cone in terms of $B$-meson distribution amplitudes (DA) defined in HQET. The on-shell $b$-quark field is replaced by the respective HQET field, and the correlation function of two-quark current is expanded in the limit of large $m_b$. Based on the observation that an increase by two units of collinear twist corresponds to a suppression by a factor of $1/m_b$, the higher twist DAs can be accounted for the power suppressed $1/m_b$ contributions in the $B$ decays \cite{Braun:2017liq}. Also, the light cone expansion is best applicable in the region of maximal recoil or small $q^2(\sim 0)$. A similar technique has been used in the extraction of the form-factors in $B\to D^{(*)}\ell\nu_{\ell}$ decays, and in the process, the c-quark mass has been kept finite, for detail see \cite{Faller:2008tr, Wang:2017jow}. The present estimate of $A_0(q^2=0)$ includes all the contributions from two and three-particle light-cone distribution amplitudes (LCDA) up to twist-four. Also, it includes the matrix elements of two-particle operators at twist-four level, which arise at next to leading (NLO) order of the light cone expansion. The twist-five and twist-six DAs are not expected to contribute to the leading power corrections $\mathcal{O}(1/m_b)$ in B-decays. As mentioned above, the power corrections up to order $1/m_b$ is known and included in the estimate of the form factors. However, the corrections at order $1/{m_b}^2$ are unknown, an error of $\sim 5\%$ due to these effects has been included in the LCSR prediction of $A_0(0)$, for detail see ref. \cite{Gubernari:2018wyi}. A sizable contribution from the two-particle states at the twist-four level has been observed. However, the twist-three and twist-four three-particle B-meson DA are shown to be insignificant numerically. Therefore, it is expected that the contributions in the sum rule from the four-particle higher-twist DA will be negligibly small since they are expected to be further suppressed by a factor $1/M^2$ where $M^2$ is the Borel parameter, and it is considered that $M^2 >> \Lambda_{QCD}^2$. Also, the NLO or higher order perturbative QCD corrections to the leading twist amplitudes are not included in the present estimate of $A_0(0)$ \cite{Gubernari:2018wyi}. It has been shown that the NLO corrections from leading-twist two particle DA are $\approx 10\%$ in the case of the form factors in $B \to D$ decays \cite{Wang:2017jow}. Other major sources of uncertainties are the threshold parameters, B-meson LCDA parameters, Borel parameter etc. Therefore, $F_2(z)$, in this case, is extracted using only the theory inputs like lattice and LCSR, without any impact from the experimental inputs on \bdstlnu. This is why all the predictions have large errors, though they are consistent with the predictions given in second and third rows of the same table. Figure \ref{fig:ddplot} compares the \bdsttaunu decay distributions for the scenarios discussed in table \ref{tab:SMpred}. We can see that the black bars and the blue bands are consistent within error bars as both the scenarios depend on exactly the same experimental inputs for the Bayesian fits. However, the results obtained using Belle 2017 data differ from these new results, as seen before.    

\section{New Physics Analysis}
\label{sec2}

\begin{table}[!htbp]
	\begin{center}
		\def\arraystretch{1.2}
		\begin{tabular}{|c|c|c|}
			\hline
			Experiment & Observable & Value\\
			\hline
			BaBar& $R(D)$& 0.440 $\pm$ 0.058  $\pm$ 0.042\\
			\cline{2-3}
			\cite{Lees:2013uzd} & $R(D^*)$ & 0.332  $\pm$ 0.024  $\pm$ 0.018 \\
			\hline
			Belle(2015) & $R(D)$ & 0.375  $\pm$ 0.064  $\pm$ 0.026 \\
			\cline{2-3}
			\cite{Huschle:2015rga} & $R(D^*)$ & 0.293  $\pm$ 0.038  $\pm$ 0.015\\
			\hline
			LHCb(2015) \cite{Aaij:2015yra} & $R(D^*)$ & 0.336  $\pm$ 0.027  $\pm$ 0.030 \\
			\hline
			Belle(2017) & $R(D^*)$ & 0.270  $\pm$ 0.035 $^{+0.028}_{-0.025}$ \\
			\cline{2-3}
			\cite{Hirose:2016wfn} & $P_{\tau}(D^*)$ & -0.38  $\pm$ 0.51 $^{+0.21}_{-0.16}$ \\
			\hline
			LHCb(2017) \cite{Aaij:2017uff}& $R(D^*)$ & 0.291  $\pm$ 0.019  $\pm$ 0.026 $\pm$ 0.013\\
			\hline
			Belle(2019) \cite{Abdesselam:2019wbt} & $F_L(D^*)$ & 0.60  $\pm$ 0.08  $\pm$ 0.04 \\
			\hline
			Belle(2019) & $R(D)$ & 0.307  $\pm$ 0.037  $\pm$ 0.016 \\
			\cline{2-3}
			\cite{Abdesselam:2019dgh} & $R(D^*)$ & 0.283  $\pm$ 0.018  $\pm$ 0.014\\
			\hline
			World averages & $R(D)$  & 0.340 $\pm$ 0.027 $\pm$ 0.013  \\
			\cline{2-3}   & $R(D^*)$  & 0.295$\pm$ 0.011 $\pm$ 0.008 \\
			\hline
		\end{tabular}
	\end{center}
	\caption{Experiment inputs for $R(D^{(*)})$, $P_{\tau}(D^*)$ and $F_L(D^*)$ used in the new physics fits.}
	\label{tab:exptinp}
\end{table}

The SM predictions given in table \ref{tab:SMpred} can be compared with the respective measurements given in table \ref{tab:exptinp}. We note that there still are discrepancies in the data on $R(D^*)$. Our predictions for $R(D^*)$ in the analysis with Belle 2019 data for \bdstlnu (with or without LCSR inputs) are consistent with the respective world average at $\sim 3\sigma$ and with the most recent result of Belle \cite{Abdesselam:2019dgh} (2019), at $\sim 1.5\sigma$. Also, the predictions of $R(D)$ \cite{hflav} are consistent with the respective world average at $\sim 1.5\sigma$. All these observations taken together could be indicating to the presence of a non-zero new physics. In this section, we will constrain possible NP scenarios from the data in a model-independent way.   

Note that the measured values of $R(D^{(*)})$, $P_{\tau}(D^*)$ and $F_L(D^*)$ are highly model-sensitive due to the model-dependence of the kinetic distribution. So, one may get different signal yields per bin from fits using different models. This could be a source of additional un-accessed systematic uncertainty in the measurements mentioned above. Consequently, the measured values obtained from fits assuming only the SM background should not be appropriate to fit the NP parameters. This problem may reduce to a large extent when the experiment will have more statistics and provide data in different $q^2$ or $w$-bins, probably in Belle-II. However, given the situation, we can still look for the possibility of large NP effects.

The most general effective Hamiltonian describing the $b\to c\tau\nu_{\tau}$ transitions is given by
\begin{align}
{\cal H}_{eff} &= \frac{4 G_F}{\sqrt{2}} |V_{cb}| \left.[( \delta_{\ell\tau} + C_{V_1}^{\ell}) {\cal O}_{V_1}^{\ell} + 
C_{V_2}^{\ell} {\cal O}_{V_2}^{\ell} + C_{S_1}^{\ell} {\cal O}_{S_1}^{\ell} \right.  \left. + C_{S_2}^{\ell} {\cal O}_{S_2}^{\ell}
+ C_{T}^{\ell} {\cal O}_{T}^{\ell}\right.]\,,
\label{eq1}
\end{align}
where  $C^\ell_W (W=V_1,V_2,S_1,S_2,T)$ are the Wilson coefficients (WCs) corresponding to the following four-Fermi operators:
\begin{align}
{\cal O}_{V_1}^{\ell} &= ({\bar c}_L \gamma^\mu b_L)({\bar \tau}_L \gamma_\mu \nu_{\ell L}), \ \ 
{\cal O}_{V_2}^{\ell} = ({\bar c}_R \gamma^\mu b_R)({\bar \tau}_L \gamma_\mu \nu_{\ell L}), \ \ 
{\cal O}_{S_1}^{\ell} = ({\bar c}_L  b_R)({\bar \tau}_R \nu_{\ell L}), \nn \\
{\cal O}_{S_2}^{\ell} &= ({\bar c}_R b_L)({\bar \tau}_R \nu_{\ell L}) , \ \ \  
{\cal O}_{T}^{\ell}   = ({\bar c}_R \sigma^{\mu\nu} b_L)({\bar \tau}_R \sigma_{\mu\nu} \nu_{\ell L})\,.
\label{eq2}
\end{align}
Here, we have considered only the left-handed neutrinos. 

\begin{table}[htbp]
\begin{center}
\def\arraystretch{1.5}
\begin{adjustbox}{width=\textwidth}
\begin{tabular}{|c|c|c|c|c|c|c|}
\hline
Case & New WCs &$R(D)$ &$R(D^*)$& $\mathcal{B}(B_c\to \tau \nu_{\tau})$ & Allowed or not? (Remarks) \\
\hline
1& $\mathcal{R}e[C_{V_1}]$ & 0.350 \errvec{11 \\ 13} &0.293 \errvec{9 \\ 10}  & 0.024 (2) & Yes \\
\hline
2& $\mathcal{R}e[C_{V_2}]$& 0.256 (16) & 0.294 \errvec{14 \\ 13} &0.024 (2) & Yes ($R(D)$ at 2$\sigma$)\\
\hline
3& $\mathcal{R}e[C_{S_1}]$&  0.377 \errvec{23 \\ 25} & 0.262 \errvec{5 \\ 6} & 0.049 (12) & Yes ($R(D^*)$ at 2$\sigma$)\\
\hline
4& $\mathcal{R}e[C_{S_2}]$&   0.324 \errvec{26 \\ 28} & 0.306 (10) &  0.877 (92) & No \\
\hline
5& $\mathcal{R}e[C_{T}]$ &  0.293 (4) & 0.303 \errvec{12 \\ 13} &  0.020 (2) & Yes ($R(D)$ at 2$\sigma$) \\
\hline
6& $\mathcal{R}e[C_{V_1}]$ ,  $\mathcal{I}m[C_{V_1}]$&  0.351 \errvec{13 \\ 12} &  0.294 \errvec{10 \\ 9} & 0.024 (2) & Yes \\
\hline
7& $\mathcal{R}e[C_{V_2}]$ , $\mathcal{I}m[C_{V_2}]$&  0.334 \errvec{32 \\ 28} & 0.298 \errvec{13 \\ 14}& 0.024 (3) & Yes\\
\hline
8& $\mathcal{R}e[C_{S_1}]$ , $\mathcal{I}m[C_{S_1}]$&  0.380 \errvec{23 \\ 25} & 0.261 (6) & 0.057 (26)  & Yes ($R(D^*)$ at 2$\sigma$) \\
\hline
9& $\mathcal{R}e[C_{S_2}]$ ,  $\mathcal{I}m[C_{S_2}]$&  0.337 \errvec{28 \\ 26} & 0.299 (11) & 0.823 (107) & No \\
\hline
10& $\mathcal{R}e[C_{T}]$ , $\mathcal{I}m[C_{T}]$&   0.296 \errvec{7 \\ 17} & 0.301 (13)  & 0.020 (2) & Yes \\
\hline
11& $\mathcal{R}e[C_{V_1}]$ ,  $\mathcal{R}e[C_{V_2}]$&  0.337 \errvec{31 \\ 30} & 0.298 \errvec{12 \\ 13} & 0.024 (3) & Yes \\
\hline
12& $\mathcal{R}e[C_{S_1}]$ , $\mathcal{R}e[C_{S_2}]$&    0.333 \errvec{29 \\ 32} & 0.299 (13) & 0.534 (224) & Marginally (see fig. \ref{fig:npplt12}) \\
\hline
\end{tabular}
\end{adjustbox}
\end{center}
\caption{ $ R(D^{(*)})$ predictions for the different new physics scenarios using the fit to all experimental inputs in table \ref{tab:exptinp}. The cases with $\mathcal{B}(B_c\to \tau \nu_{\tau}) > 30\%$ are physically ruled out.}
\label{tab:npfitresBa}
\end{table}

\begin{table}[htbp]
\begin{center}
\def\arraystretch{1.5}
\begin{adjustbox}{width=\textwidth}
\begin{tabular}{|c|c|c|c|c|c|c|}
\hline
Case & New WCs &$R(D)$ &$R(D^*)$& $\mathcal{B}(B_c\to \tau \nu_{\tau})$ & Allowed or not? (Remarks) \\
\hline
1& $\mathcal{R}e[C_{V_1}]$ & 0.338 \errvec{13 \\ 14} &0.283 (10) & 0.023 (2) & Yes \\
\hline
2& $\mathcal{R}e[C_{V_2}]$& 0.261 \errvec{17 \\ 18} & 0.288 \errvec{16 \\ 15} & 0.023 (3) & Yes ($R(D)$ at 2$\sigma$)  \\
\hline
3& $\mathcal{R}e[C_{S_1}]$&  0.355 \errvec{26 \\ 27} & 0.259 \errvec{5 \\ 6}  & 0.039 (12) &
Yes ($R(D^*)$ at 2$\sigma$) \\
\hline
4& $\mathcal{R}e[C_{S_2}]$&   0.301 (30)  & 0.302 (10)  & 0.826 (95)
& No \\
\hline
5& $\mathcal{R}e[C_{T}]$ &  0.294 (4) & 0.292 \errvec{14 \\ 15}  & 0.020 (2) & Yes ($R(D)$ at 2$\sigma$) \\
\hline
6& $\mathcal{R}e[C_{V_1}]$ ,  $\mathcal{I}m[C_{V_1}]$&  0.339 \errvec{14 \\ 13} &  0.285 \errvec{12 \\ 9} & 0.023 (2) & Yes  \\
\hline
7& $\mathcal{R}e[C_{V_2}]$ , $\mathcal{I}m[C_{V_2}]$&  0.307 \errvec{30 \\ 32} & 0.292 \errvec{15 \\ 14} & 0.024 (3) & Yes  \\
\hline
8& $\mathcal{R}e[C_{S_1}]$ , $\mathcal{I}m[C_{S_1}]$&  0.357 \errvec{25 \\ 28} & 0.257 (6) & 0.049 (33) & Yes ($R(D^*)$ at 2$\sigma$) \\
\hline
9& $\mathcal{R}e[C_{S_2}]$ ,  $\mathcal{I}m[C_{S_2}]$&  0.316 \errvec{31 \\ 29}  & 0.294 (13)  & 0.755 (71) &  No \\
\hline
10& $\mathcal{R}e[C_{T}]$ , $\mathcal{I}m[C_{T}]$&   0.298 \errvec{7 \\ 12} & 0.292 \errvec{15 \\ 13} & 0.020 (2) & Yes \\
\hline
11& $\mathcal{R}e[C_{V_1}]$ ,  $\mathcal{R}e[C_{V_2}]$&  0.318 \errvec{31 \\ 29} & 0.291 \errvec{15 \\ 14} & 0.024 (3) & Yes  \\
\hline
12& $\mathcal{R}e[C_{S_1}]$ , $\mathcal{R}e[C_{S_2}]$&   0.316 \errvec{31 \\ 29} & 0.292 (14) & 0.409 (228) & Marginally (see fig. \ref{fig:npplt12}) \\
\hline
\end{tabular}
\end{adjustbox}
\end{center}
\caption{ $ R(D^{(*)})$ predictions for the different new physics scenarios using the fit to all experimental inputs (in table \ref{tab:exptinp}) except the ones from BaBar. The cases with $\mathcal{B}(B_c\to \tau \nu_{\tau}) > 30\%$ are physically ruled out.}
\label{tab:npfitres}
\end{table}

\begin{figure}[htbp]
	\begin{center}
		\subfigure[Case 1]{%
			\label{fig:nppltpdf1}
			\includegraphics[width=0.45\textwidth]{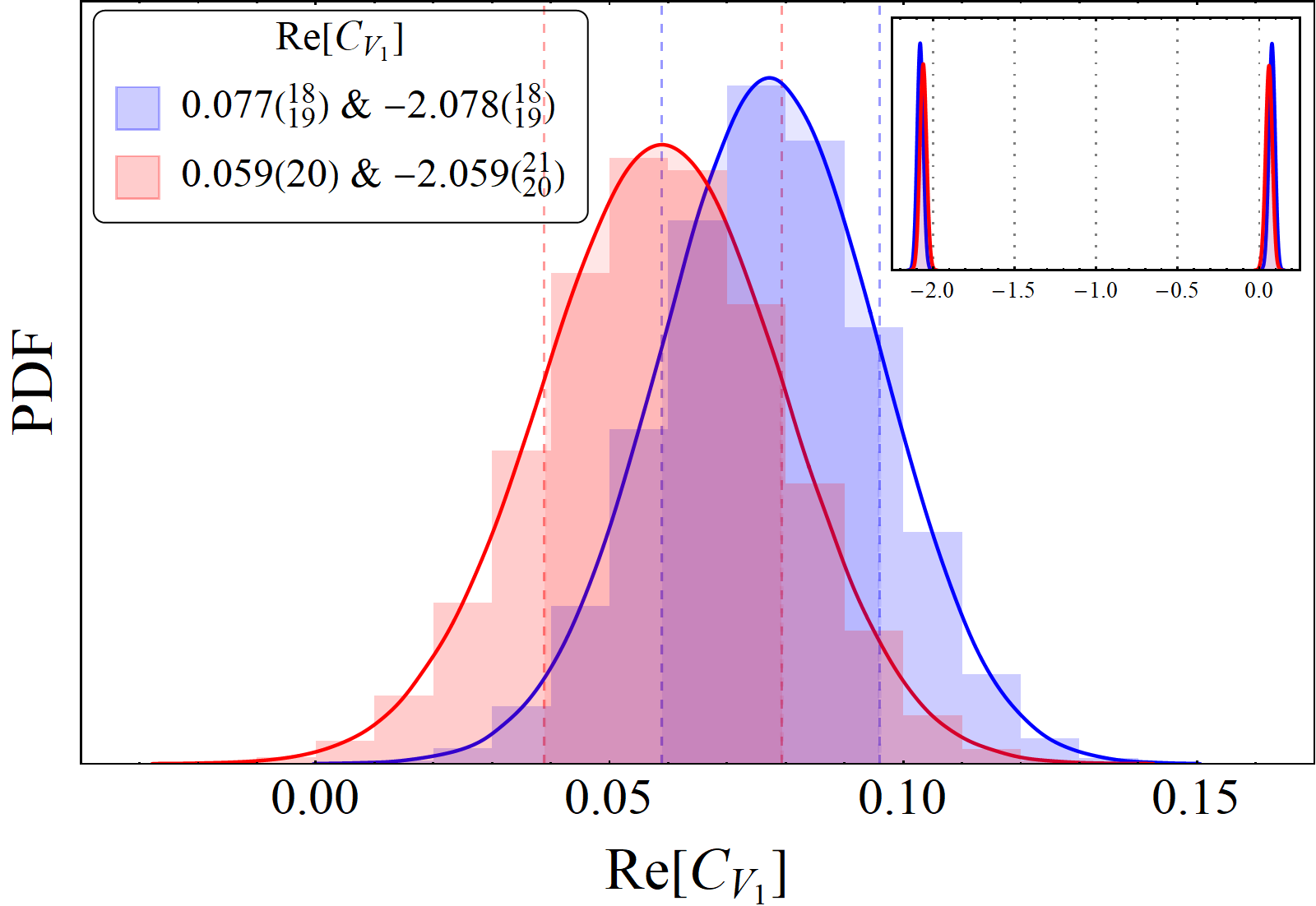}
		}%
		~~~~
		\subfigure[Case 2]{%
			\label{fig:nppltpdf2}
			\includegraphics[width=0.45\textwidth]{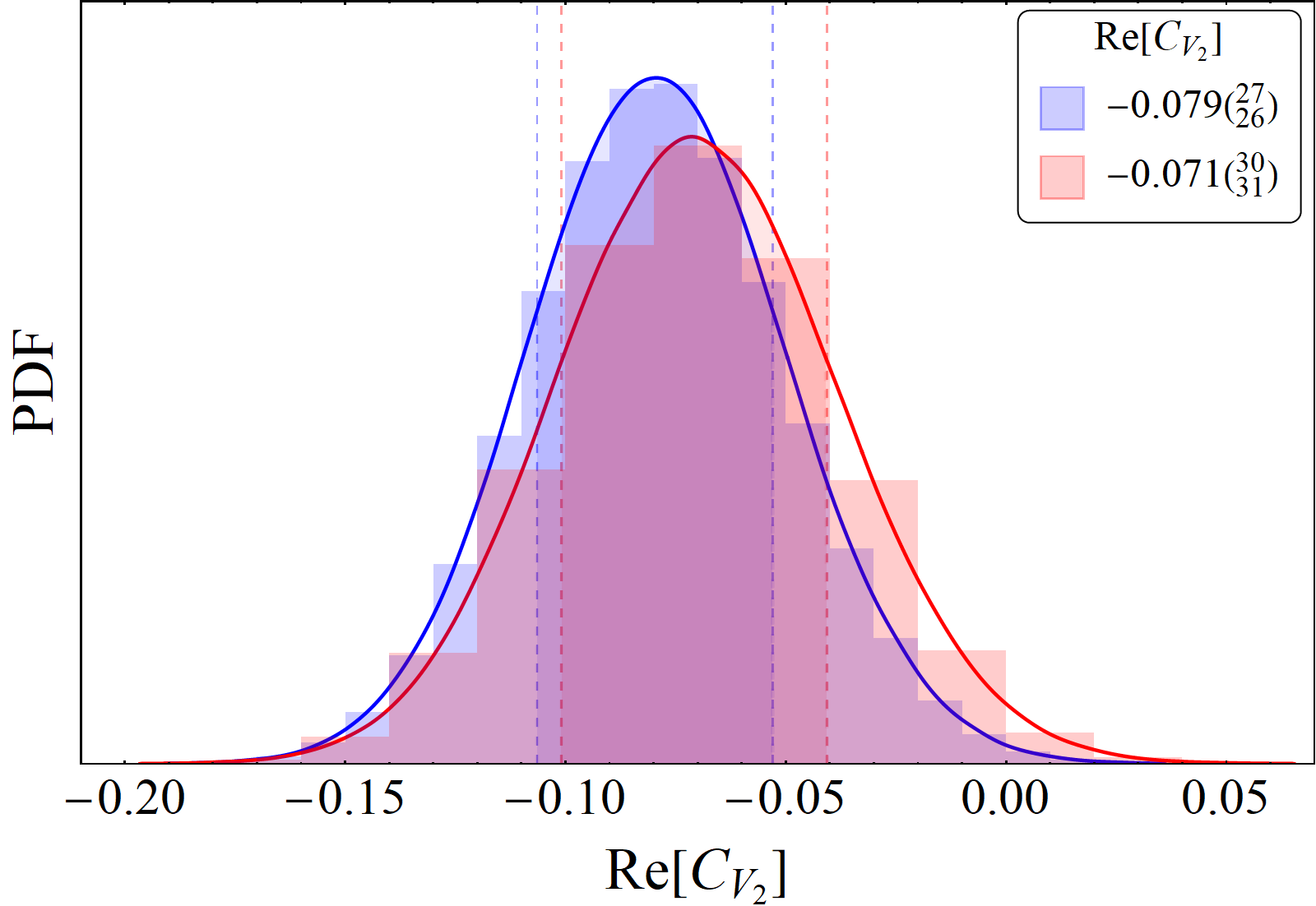}
		}%
		\\
		\subfigure[Case 3]{%
			\label{fig:nppltpdf3}
			\includegraphics[width=0.45\textwidth]{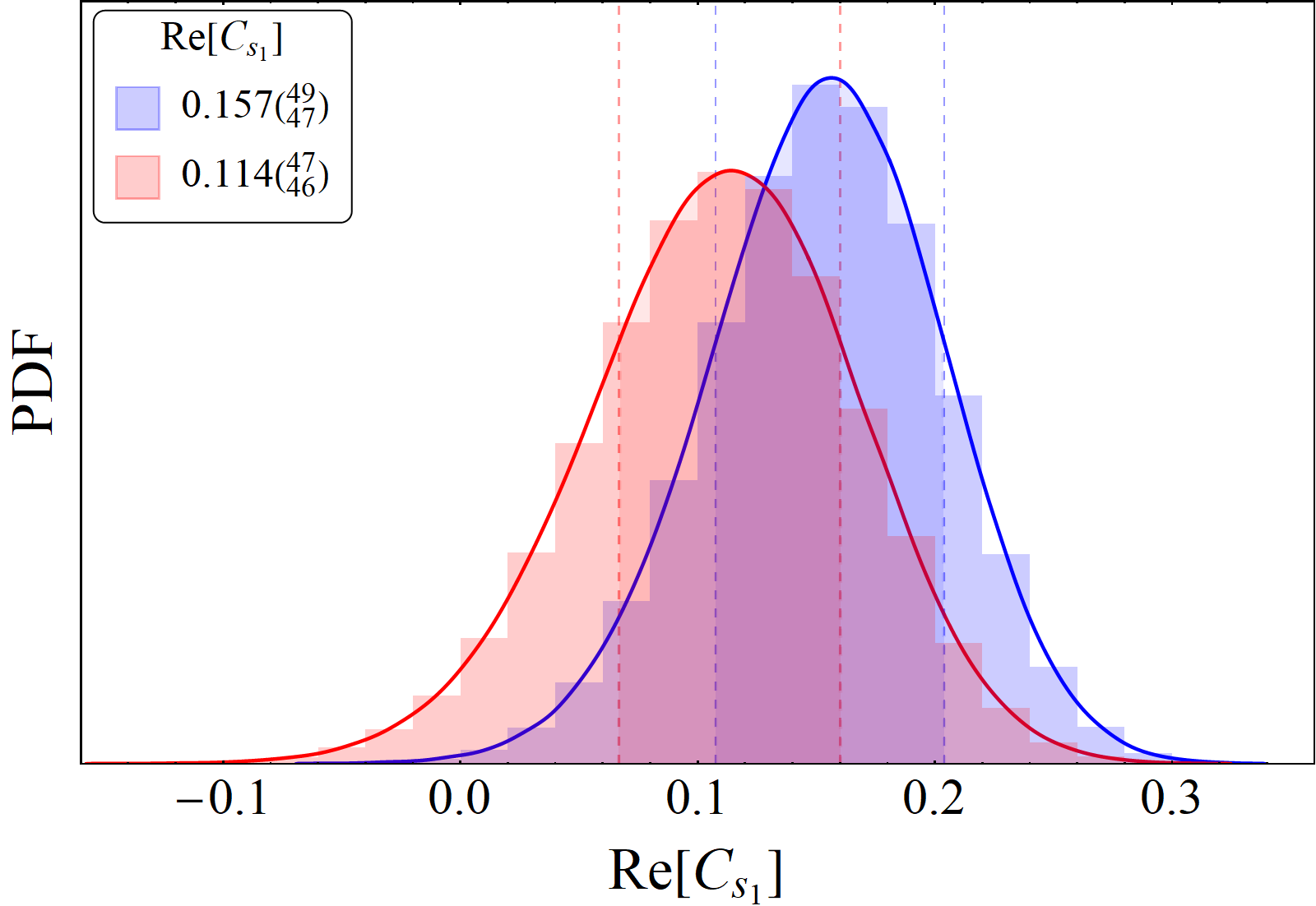}
		}%
		~~~~
		\subfigure[Case 5]{%
			\label{fig:nppltpdf5}
			\includegraphics[width=0.464\textwidth]{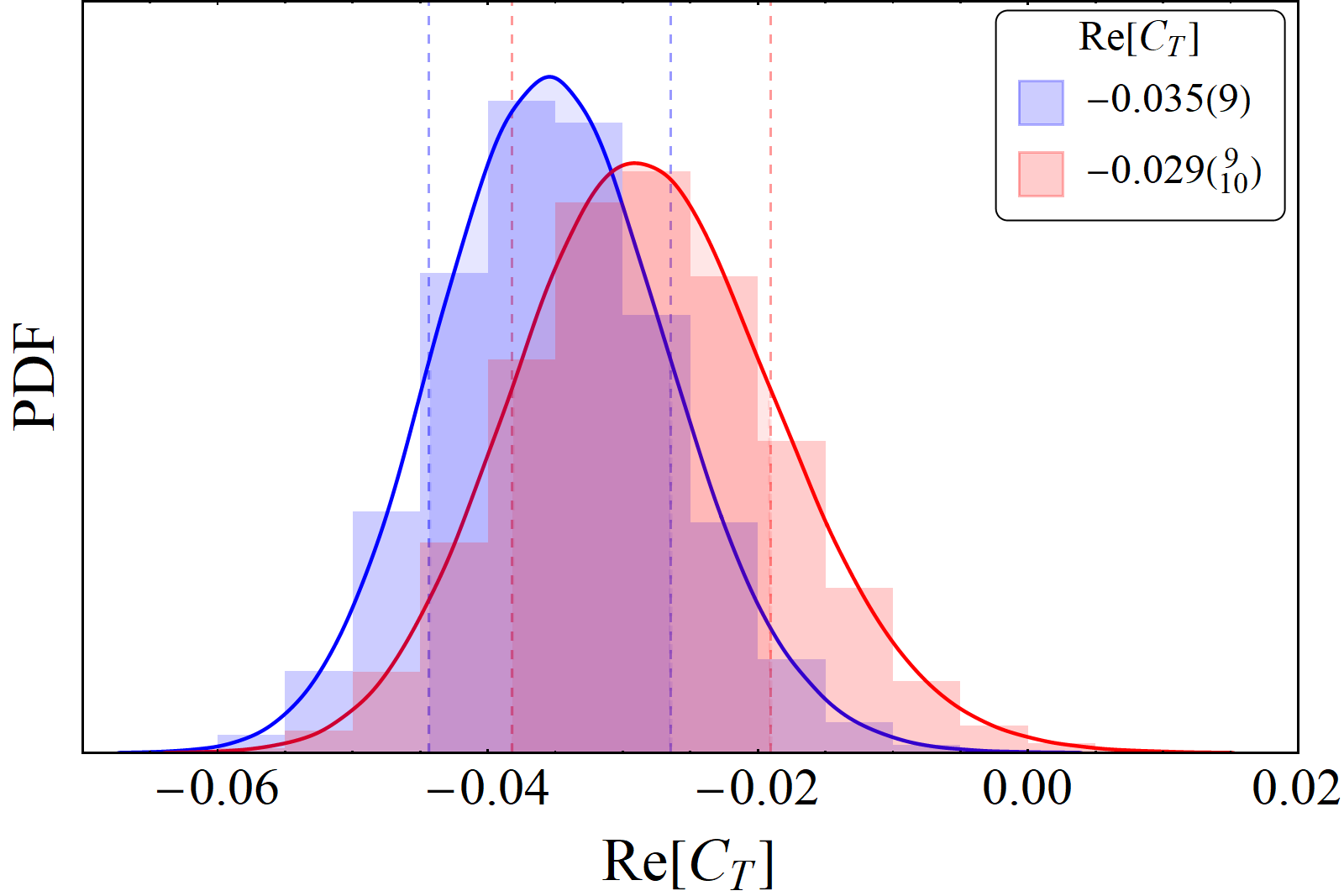}
		}%
	\end{center}
	\caption{1D marginal posterior distributions for the NP WCs for the allowed one parameter scenarios in tables \ref{tab:npfitresBa} and \ref{tab:npfitres}. The blue ones are obtained with all the experimental inputs in table \ref{tab:exptinp} while for the red ones, we dropped the inputs from BaBar. The region between the dashed lines contain $68\%$ credible intervals (high density region around mode) and the corresponding values are also listed. The marginal for $\mathcal{R}e[C_{V_1}]$ (Case 1) has two distinct modes (shown inset fig.\ref{fig:nppltpdf1}) but we focus on the mode corresponding to lowest absolute value for the WC.}
	\label{fig:posteriors}
\end{figure}

\begin{figure}[htbp]
	\begin{center}
		\subfigure[Case 6]{%
			\label{fig:npplt6}
			\includegraphics[width=0.3\textwidth]{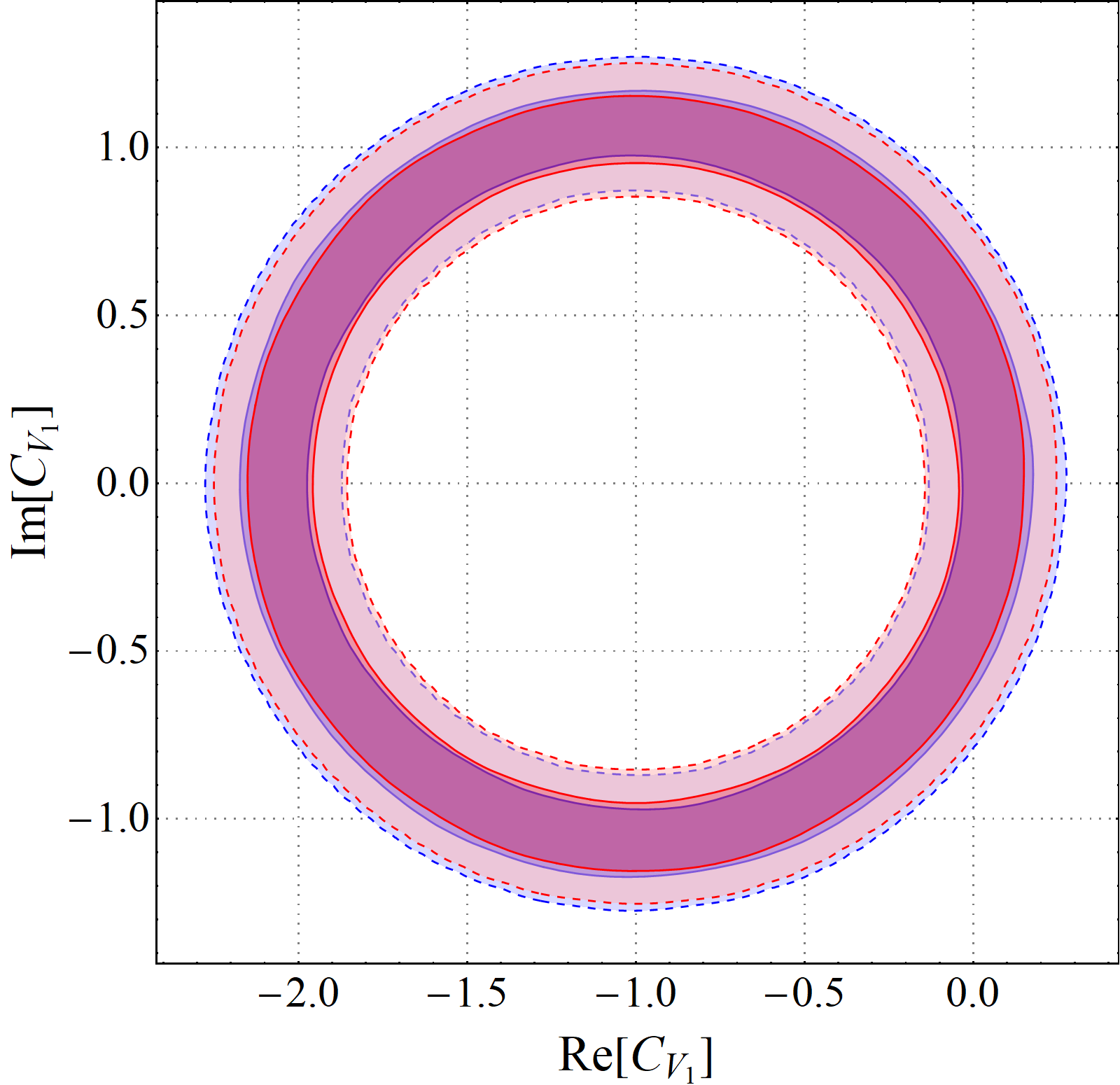}
		}%
		~~
		\subfigure[Case 7]{%
			\label{fig:npplt7}
			\includegraphics[width=0.304\textwidth]{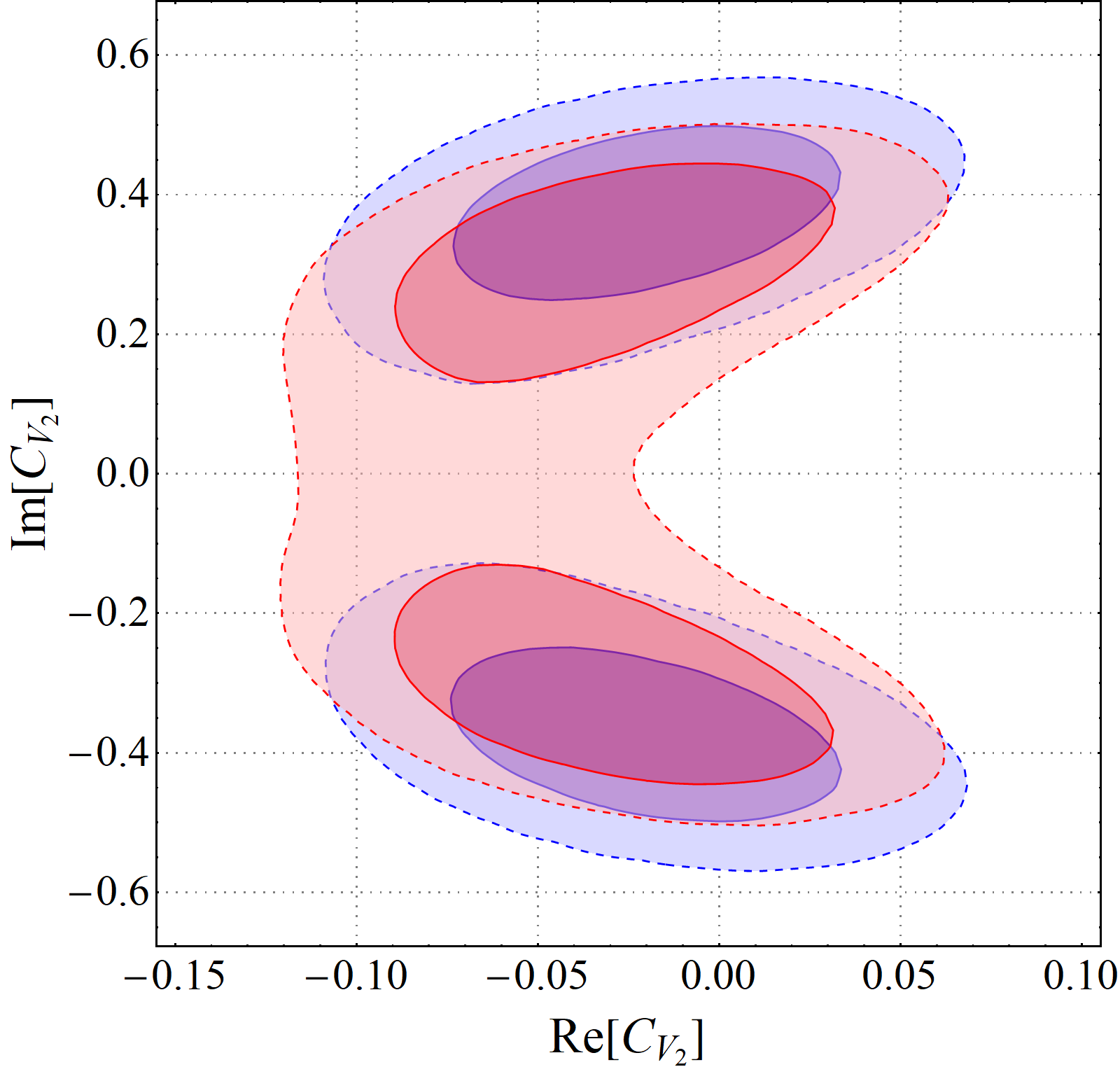}
		}%
		~~
		\subfigure[Case 8]{%
			\label{fig:npplt8}
			\includegraphics[width=0.3\textwidth]{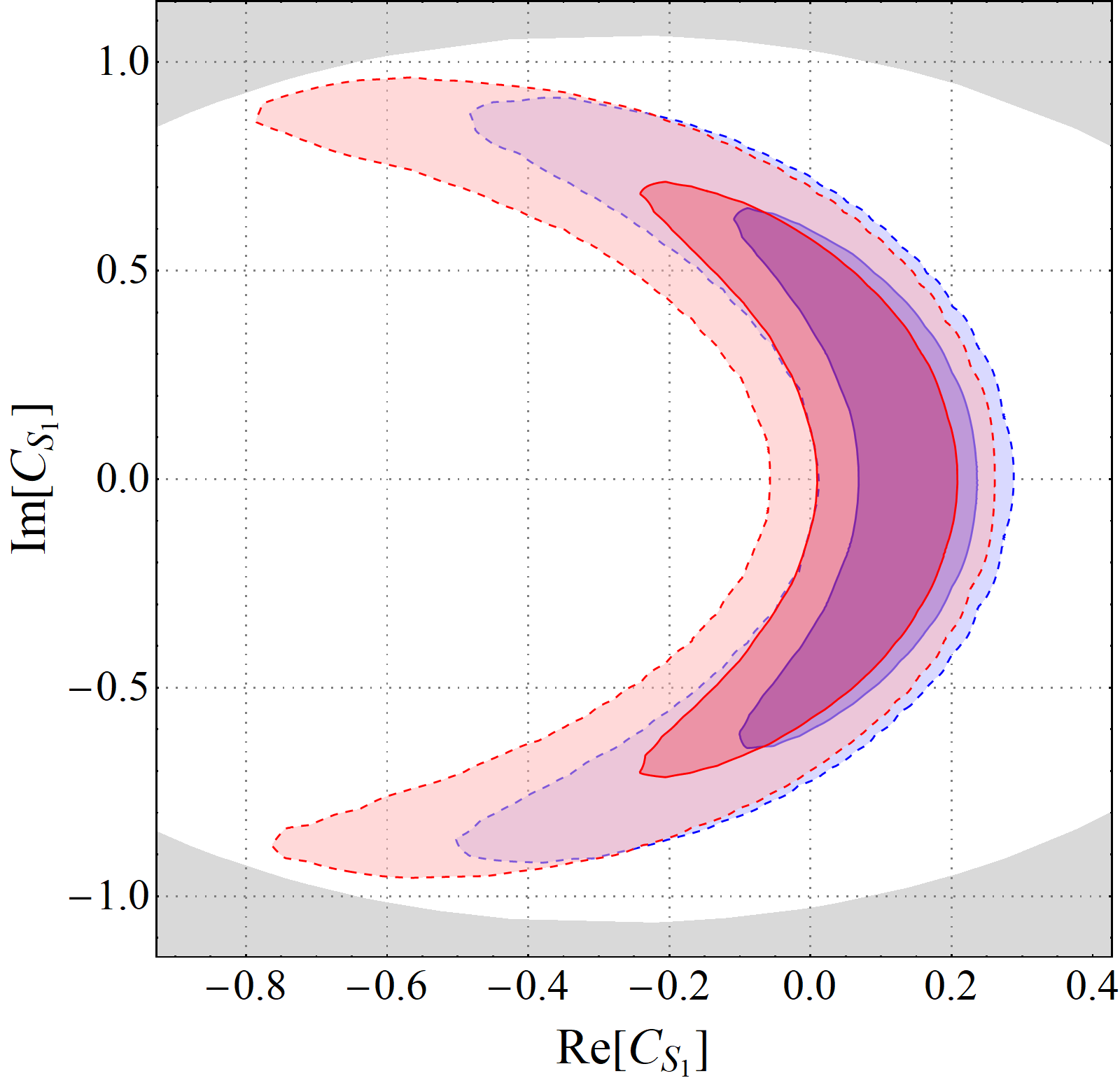}
		}%
		\\
		\subfigure[Case 9]{%
			\label{fig:npplt9}
			\includegraphics[width=0.3\textwidth]{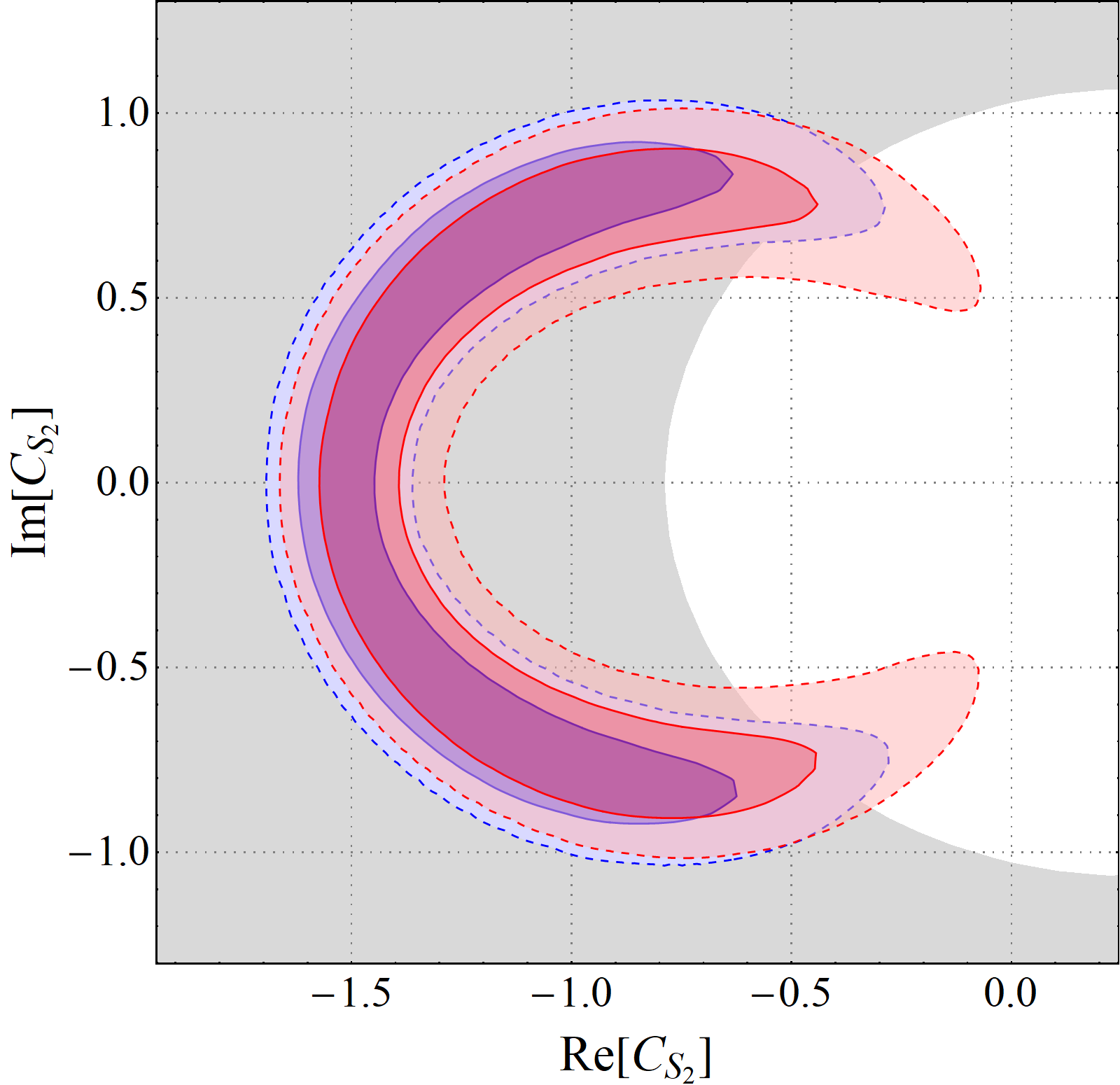}
		}%
	    ~~
	    \subfigure[Case 10]{%
	    	\label{fig:npplt10}
	    	\includegraphics[width=0.301\textwidth]{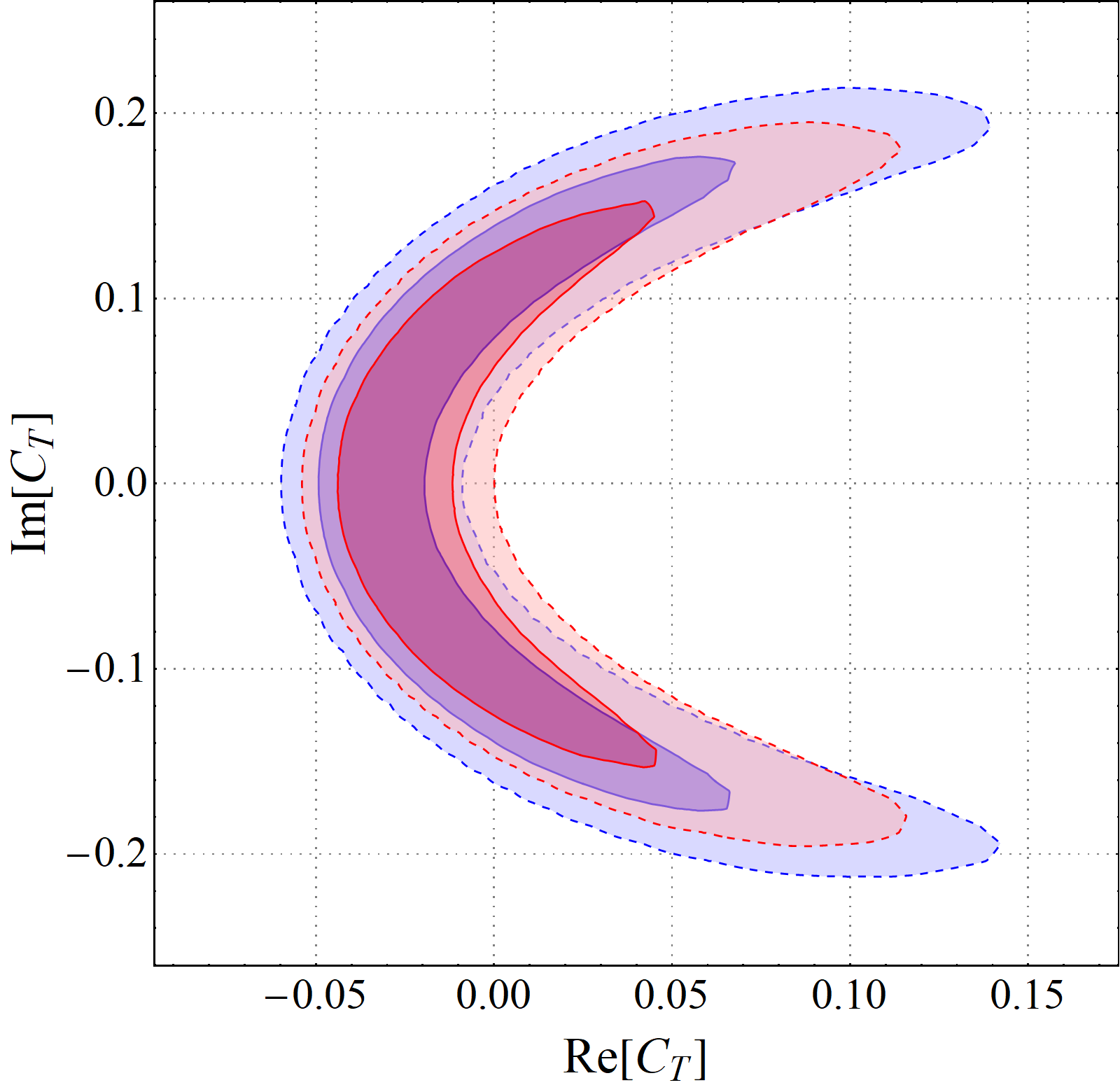}
	    }%
	    ~~
	    \subfigure[Case 11]{%
	    	\label{fig:npplt11}
	    	\includegraphics[width=0.3\textwidth]{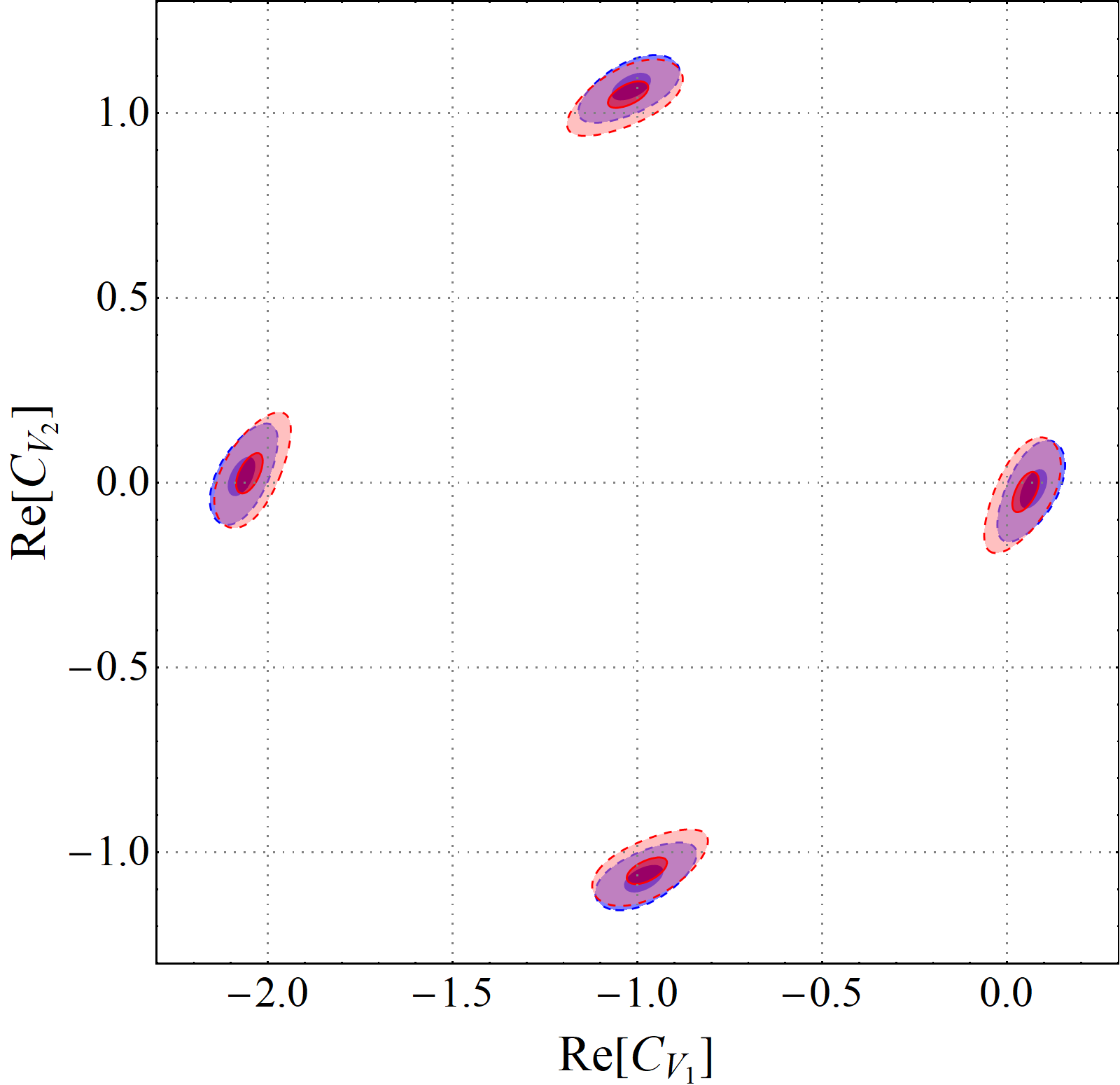}
	    }%
	    \\
	    \subfigure[Case 12]{%
	    	\label{fig:npplt12}
	    	\includegraphics[width=0.3\textwidth]{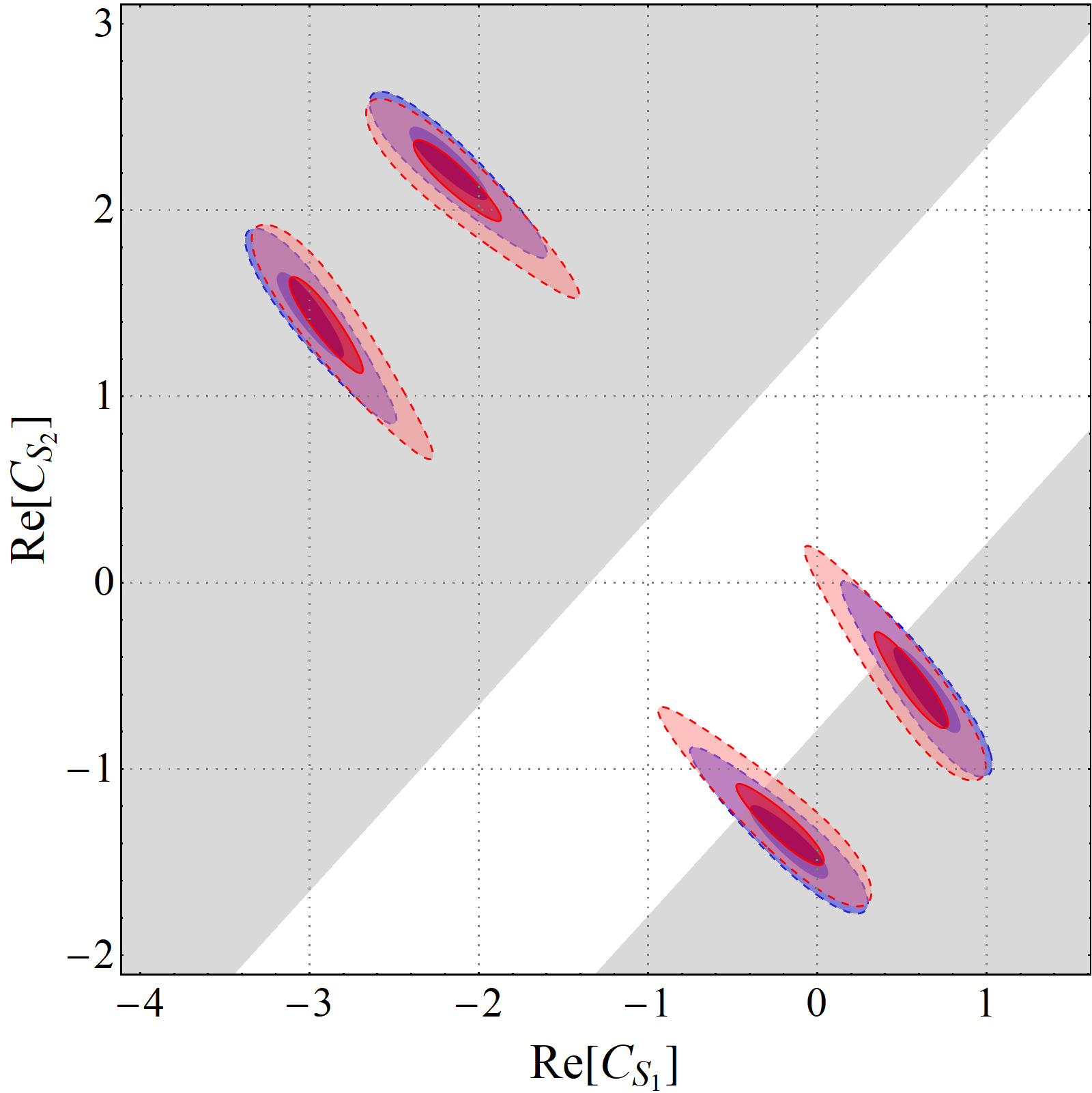}
	    }%
	\end{center}
	\caption{2D marginal posterior distributions for two parameter NP scenarios in tables \ref{tab:npfitresBa} and \ref{tab:npfitres}. Blue shaded regions correspond to the NP fits to all the experimental inputs in table \ref{tab:exptinp} while the red shaded regions correspond to the fits with BaBar data dropped. Solid and dashed contours enclose respectively $68\%$ and  $95\%$ highest probability regions for figures \ref{fig:npplt6} - \ref{fig:npplt10}, and $1$ and $4\sigma$ CLs for figures \ref{fig:npplt11} and \ref{fig:npplt12}. The gray shaded regions denote the NP parameter space disallowed by $\mathcal{B}(B_c\to \tau \nu_{\tau}) > 30\%$.}
	\label{fig:correlations}
\end{figure}

\begin{figure}[htbp]
	\begin{center}
		\subfigure[Case 11]{%
			\label{fig:npplt11comb}
			\includegraphics[width=0.465\textwidth]{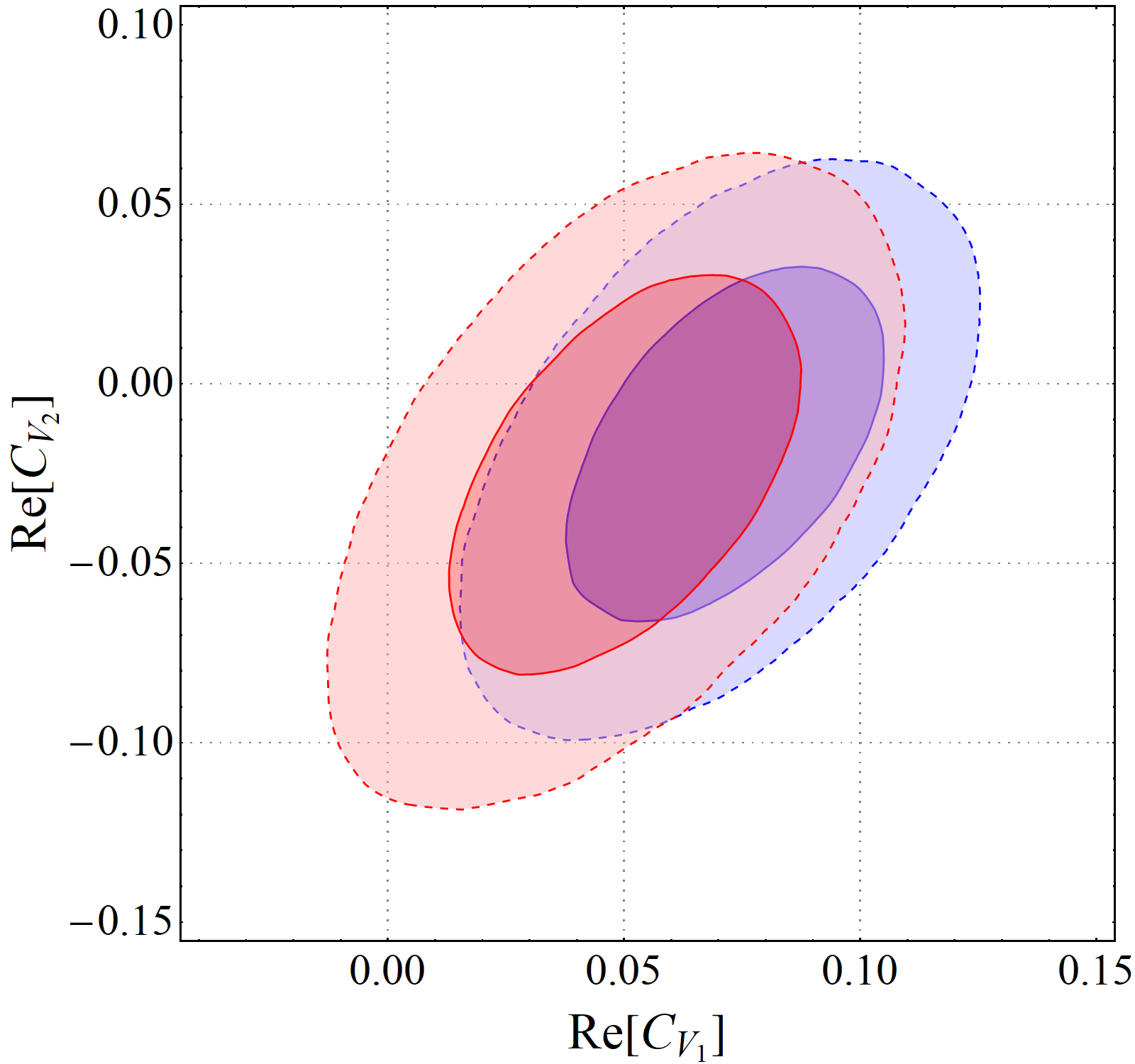}
		}%
		~~~~~
		\subfigure[Case 12]{%
			\label{fig:npplt12comb}
			\includegraphics[width=0.45\textwidth]{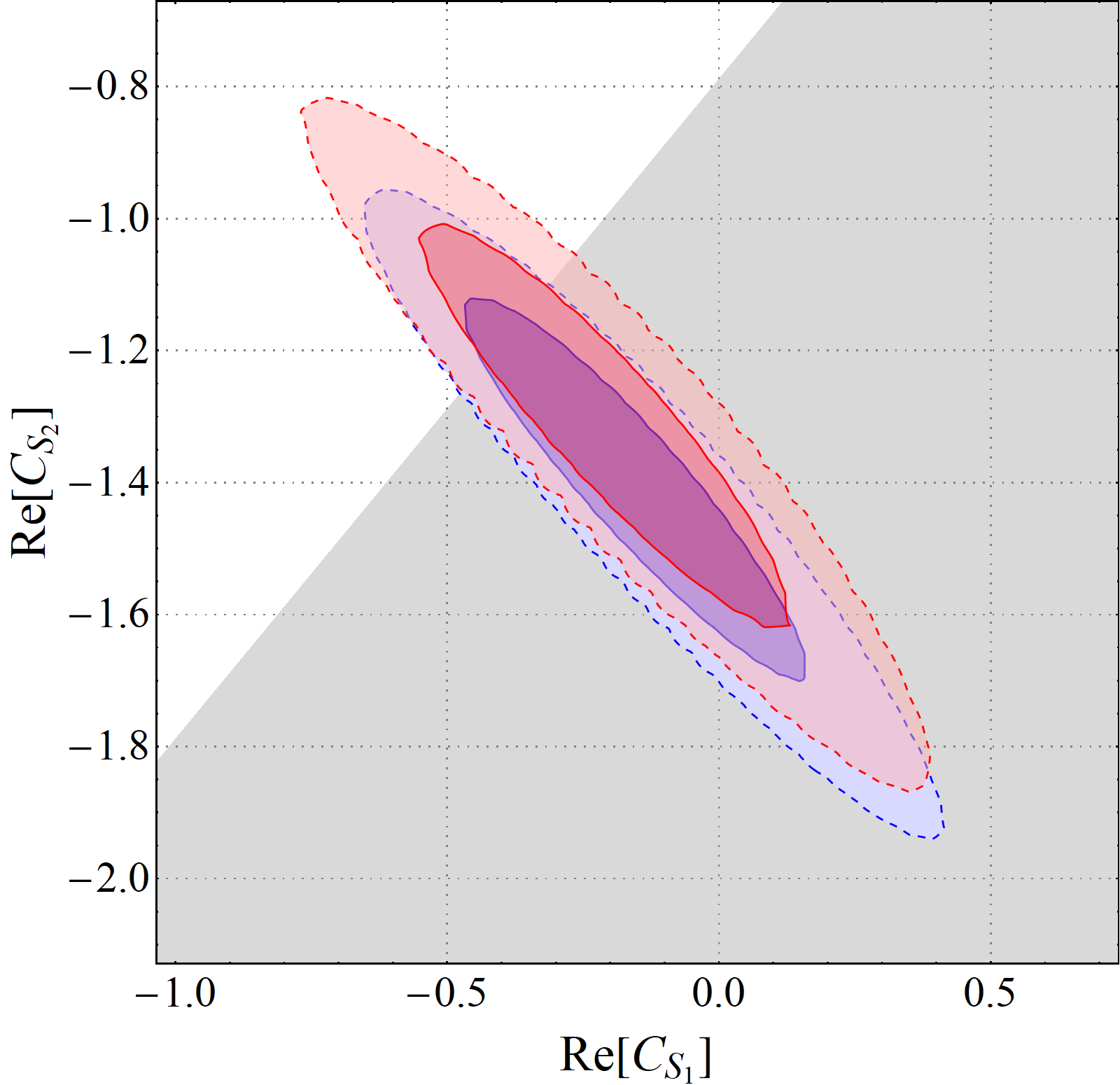}
		}%
	\end{center}
	\caption{ Allowed NP parameter space for the multi-modal new physics scenarios, with the smallest absolute values for the WCs. The blue regions correspond to the NP fits to all the experimental inputs in table \ref{tab:exptinp} while the red ones correspond to the fits with BaBar data dropped. Solid and dashed contours enclose $68\%$ and $95\%$ highest probability regions respectively. The gray shaded regions denote the NP parameter space disallowed by $\mathcal{B}(B_c\to \tau \nu_{\tau}) > 30\%$.}
	\label{fig:correlationscomb}
\end{figure}

\begin{table}[t]
\begin{center}
\def\arraystretch{1.5}
\begin{adjustbox}{width=\textwidth}
\begin{tabular}{|c|c|c|c|c|c|c|c|}
\hline
Case & New WCs &$P_{\tau}(D)$ &$A_{FB}(D)$&$P_{\tau}(D^*)$ & $F_L(D^*) $& $A_{FB}(D^*)$ \\
\hline
1& $\mathcal{R}e[C_{V_1}]$ & 0.326 (3) &0.3596 (4)   &-0.487 (24) & 0.471 (10) & -0.035 \errvec{20 \\ 24} \\
\hline
2& $\mathcal{R}e[C_{V_2}]$&0.326 (3)& 0.3596 (4)&-0.164 \errvec{31 \\ 30} & 0.482 (10)  & 0.0001  \errvec{203 \\ 200} \\
\hline
3& $\mathcal{R}e[C_{S_1}]$&  0.464  \errvec{35 \\ 33} & 0.3597 \errvec{4 \\ 3} &-0.432 \errvec{29 \\ 31} & 0.492 \errvec{11 \\ 12} & -0.004 \errvec{23 \\ 21} \\
\hline
5& $\mathcal{R}e[C_{T}]$ &  0.348 (6) & 0.344 (4) &-0.455 (20)  & 0.453 \errvec{9 \\ 11} & 0.008 \errvec{18 \\ 17} \\
\hline
6& $\mathcal{R}e[C_{V_1}]$ ,  $\mathcal{I}m[C_{V_1}]$&  0.325 \errvec{2 \\ 3} &  0.3597  \errvec{4 \\ 3} &-0.488 \errvec{24 \\ 25} & 0.474 \errvec{13 \\ 8} & -0.032 \errvec{24 \\ 20} \\
\hline
7& $\mathcal{R}e[C_{V_2}]$ , $\mathcal{I}m[C_{V_2}]$&  0.326 \errvec{3 \\ 2} & 0.3596  \errvec{3 \\ 4}&-0.184 \errvec{32 \\ 29} & 0.473 \errvec{11 \\ 10} & 0.012 \errvec{18 \\ 19} \\
\hline
8& $\mathcal{R}e[C_{S_1}]$ , $\mathcal{I}m[C_{S_1}]$&  0.471 \errvec{36 \\ 32} & 0.3597 \errvec{4 \\ 3} &-0.439 \errvec{32 \\ 27} & 0.487 \errvec{12 \\ 14} & -0.014 \errvec{22 \\ 25} \\
\hline
10& $\mathcal{R}e[C_{T}]$ , $\mathcal{I}m[C_{T}]$&   0.343  \errvec{32 \\ 10} & 0.348 \errvec{6 \\ 17} &-0.437 \errvec{31 \\ 69} & 0.445 \errvec{34 \\ 16} & 0.014 \errvec{20 \\ 17} \\
\hline
11& $\mathcal{R}e[C_{V_1}]$ ,  $\mathcal{R}e[C_{V_2}]$&  0.326 (3)  & 0.3597 \errvec{4 \\ 3} &-0.490 \errvec{24 \\ 25} & 0.472 \errvec{10 \\ 11} & 0.303 \errvec{21 \\ 20} \\
\hline

\end{tabular}
\end{adjustbox}
\end{center}
\caption{  Predictions for various angular observables for the allowed new physics scenarios in table \ref{tab:npfitresBa} using the NP fit to all the experimental inputs.}
\label{tab:npfitresangobsBa}
\end{table}

We fit these Wilson coefficients to the available data on the integrated observables in \bdsttaunu from BaBar, Belle and LHCb collaborations in order to estimate the size of the different NP effects allowed by the present experimental scenario. All the different experimental inputs to this NP fit are listed in table \ref{tab:exptinp}. Detailed expressions for the observables $R(D^*)$, $P_{\tau}(D^*)$, $F_L(D^*)$, and $\mathcal{B}(B_c\to \tau \nu_{\tau})$ in terms of the WCs can be found in the references \cite{Sakaki:2013bfa,Alonso:2016oyd}. We consider several simple NP scenarios with both real and complex WCs, with a maximum of two NP fit parameters at a time, the respective cases are listed in tables \ref{tab:npfitresBa} and \ref{tab:npfitres}. In these Bayesian fits, the results of the analysis without LCSR inputs in Section \ref{sec1} are used to define the prior distributions for the BGL coefficients for $N=2$ for all the \bdstlnu form-factors while for the prior distributions for $B \to D$ form-factors, we depend on the \bdlnu analysis in our earlier work \cite{Jaiswal:2017rve}.

Table \ref{tab:npfitresBa} gives the predictions of $R(D^{(*)})$ for different NP scenarios obtained from a fit to all the experimental inputs while table \ref{tab:npfitres} lists the $ R(D^{(*)})$ predictions corresponding to the fit with experimental inputs from BaBar dropped. As was already pointed out in reference \cite{Alonso:2016oyd}, the cases with $\mathcal{B}(B_c\to \tau \nu_{\tau}) > 30\%$ are physically ruled out. In the SM, we get a prediction for $\mathcal{B}(B_c\to \tau \nu_{\tau}) = 0.020 \pm 0.002$. The tensor operators do not contribute to the  transition $B_c\to \tau \nu_{\tau}$, and hence the NP parameter space for WC `$C_{T}$' cannot be constrained by the bounds on $\mathcal{B}(B_c\to \tau \nu_{\tau})$. We note that apart from the scenario with the WC `$C_{S_2}$' (both real and complex), all other scenarios are favored (strongly, more or less) by the present data.  Among the most favored NP scenarios, there are SM-type new four-fermion operators (${\cal O}_{V_1}$) with real or complex WCs. 

Marginal posterior distributions of the WCs for all the allowed one parameter scenarios are shown in figure \ref{fig:posteriors} with the $1\sigma$ Credible Intervals (CI) around the central moment of the relevant WCs mentioned. The blue regions correspond to the NP fits to all the experimental inputs in table \ref{tab:exptinp}, while the red ones correspond to those with the experimental inputs from BaBar dropped. Note that, in some cases, there are slight changes in the allowed regions after dropping the BaBar data. For the two parameter scenarios, we have shown the correlations between the WCs in figures \ref{fig:correlations} and \ref{fig:correlationscomb}. The solid and the dashed contours (in figures \ref{fig:npplt6} - \ref{fig:npplt10} and \ref{fig:correlationscomb})  enclose the $68\%$ and  $95\%$ probability regions, respectively. The gray shaded regions depict the NP parameter space disallowed by the constraint $\mathcal{B}(B_c\to \tau \nu_{\tau}) > 30\%$. Note that the two-operator scenario with $Re(C_{S_1})$ and $Re(C_{S_2})$ (Case 12) is only marginally allowed as most of the parameter space, allowed by \bdsttaunu and \bdtaunu data, are disfavored by the limit on $\mathcal{B}(B_c\to \tau \nu_{\tau})$. The 2D marginal posterior distributions for cases 7, 11, and 12 are found to be multi-modal i.e., there are multiple allowed regions in the NP parameter space.

A note here: as can be clearly seen from the plots, the multiple modes of cases 11 and 12 are fairly disconnected and separated from each other. As a result, the Markov-Chain-Monte-Carlo (MCMC) process, which samples the parametric distributions, has a fair chance to get stuck at one of these modes. Though this can be avoided using some parallel-tempering method \cite{PhysRevLett.57.2607} or an affine-invariant ensemble sampler \cite{goodman2010}, we instead have chosen to show the $1$ and $4 \sigma$ confidence intervals obtained from the frequentist analysis in figures \ref{fig:npplt11} and \ref{fig:npplt12}. On the other hand, figure \ref{fig:correlationscomb} only shows the $68\%$ and $95\%$ Bayesian credible intervals for the mode that corresponds to the smallest absolute values for the WCs for the cases 11 and 12. As can be seen from these plots, the regions obtained from frequentist and Bayesian analyses are consistent. The nature of correlations are found to be mostly consistent with those in ref. \cite{Bhattacharya:2018kig}, which does a similar NP analysis with CLN parametrization using the old Belle data \cite{Abdesselam:2017kjf}. These results could be utilized to extract the couplings and masses of different NP models.

In table \ref{tab:npfitresangobsBa}, we have shown the predicted values of the angular observables obtained from the fit to all $B \to D^{(*)} \tau \nu_{\tau}$ data (in table \ref{tab:exptinp}) for all the allowed NP scenarios (in table \ref{tab:npfitresBa}). If we instead use the fit with both experimental and lattice inputs from \bdlnu decay (for details, see ref. \cite{Jaiswal:2017rve}), we get the Standard Model estimates : $P_{\tau}(D) = 0.326(3)$ and $A_{FB}(D) = 0.3596(3)$. Note that the predicted values of $P_{\tau}(D)$ are different (slightly higher) in the scenarios with left-handed scalar current (i.e., with WC `$C_{S_1}$') than the other scenarios. Similar observation can be made for $P_{\tau}(D^*)$ in the scenario with right-handed vector current (i.e., with WC `$C_{V_2}$'). In this scenario, the magnitude of the predicted value is much lower than those of the predictions in other scenarios. Therefore, the precise measurements of these observables will be helpful to pinpoint these specific NP scenarios. Also, while the predicted values of $F_L(D^*)$ in all the NP scenarios are consistent with the experimental result at 2$\sigma$, their central values are much lower than the corresponding measured value \cite{Abdesselam:2019wbt}.

\section{$R(D^*)$ using the preliminary lattice inputs at non-zero recoils}
\label{subsec:lattice}

\begin{table}[htbp]
	\begin{center}
	\def\arraystretch{1.3}
	\begin{tabular}{|c|c|c|c|}
		\hline
		Parameters &Fit to JLQCD data\cite{Kaneko:2019vkx}  & Fit to MILC data\cite{Aviles-Casco:2019zop}\\
		& + LCSR data \cite{Gubernari:2018wyi} & + LCSR data \cite{Gubernari:2018wyi} \\
		 & + $h_{A_1}(1)$ from MILC \cite{Bailey:2014tva}&  + $h_{A_1}(1)$ from MILC \cite{Bailey:2014tva}\\
		\hline
		$a^f_0$ & 0.0120 (1) & 0.0125 (1) \\
		$a^f_1$  & -0.0084 \errvec{144 \\ 143}& 0.0087 (186)\\
		$a^f_2$   &  0.1730 (4082)  & -0.2684 (3802) \\
		\hline
		$a^{F_1}_1$&  -0.0015 \errvec{30 \\ 31}  & -0.0040 \errvec{32 \\ 34} \\
		$a^{F_1}_2$ & 0.0885 \errvec{638 \\ 566}& 0.0769 \errvec{708 \\ 750}\\
		\hline
		$a^g_0$ &  0.0299 (4) & 0.0329 (6)\\
		$a^g_1$  & -0.0634 \errvec{481 \\ 476} & -0.1407 \errvec{826 \\ 784} \\
		$a^g_2$ &  -0.0285 (5667) & 0.0173 (5658) \\
		\hline
       $a^{F_2}_0$  & 0.0478 (19) & 0.0505 (12) \\
      $a^{F_2}_2$ &  0.0192 (5674) & -0.0072 (5646) \\
      \hline
	\end{tabular}
	\end{center}
\caption{The values of the BGL coefficients (N=2) extracted using the preliminary lattice results on form-factors beyond zero-recoil from MILC collaboration \cite{Aviles-Casco:2019zop} and JLQCD \cite{Kaneko:2019vkx}. In both the analyses, LCSR inputs\cite{Gubernari:2018wyi} for $q^2=0$ are used and  $h_{A_1}(1) = 0.906(13)$ is taken from unquenched Fermilab/MILC lattice data \cite{Bailey:2014tva}.}
	\label{tab:BGLfitlat}
\end{table}

   \begin{table}[htbp]
   	\begin{center}
   		\def\arraystretch{1.5}
   		\begin{adjustbox}{width=\textwidth}
   			\begin{tabular}{|c|cccc|cccc|}
   				\hline
   				Inputs used&\multicolumn{4}{c|}{ JLQCD data\cite{Kaneko:2019vkx} + LCSR data \cite{Gubernari:2018wyi}}&\multicolumn{4}{c|}{MILC data\cite{Aviles-Casco:2019zop} + LCSR data \cite{Gubernari:2018wyi}}\\
   				in the fit &\multicolumn{4}{c|}{+ $h_{A_1}(1)$ from MILC \cite{Bailey:2014tva}} &\multicolumn{4}{c|}{+ $h_{A_1}(1)$ from MILC \cite{Bailey:2014tva}}\\
   				\hline
   				Observable&$R({D^*})$&$P_{\tau}(D^*)$&$F_L(D^*)$&$A_{FB}(D^*)$&$R({D^*})$&$P_{\tau}(D^*)$&$F_L(D^*)$&$A_{FB}(D^*)$\\
   				\hline
   				Value & 0.244(12)&-0.500\errvec{14 \\ 12}&0.458(11)&-0.051\errvec{12 \\ 11}  &0.251(12)& -0.507\errvec{10 \\ 13}&0.443(10)&-0.062\errvec{12\\11}  \\
   				\hline
   				& 	1&-0.885&-0.792&-0.780     &1&-0.911&-0.854&-0.739  \\
   				Correlation &&1&0.956&0.917          &&1&0.959&0.834  \\
   				Matrix & 	&&1&0.954               &&&1&0.911  \\
   				& 	&&&1                &&&&1 \\
   				\hline
   			\end{tabular}
   		\end{adjustbox}
   	\end{center}
   	\caption{Predictions for and the correlations between the observables in $B\to D^{*}\tau \nu_{\tau}$ decays corresponding to the fits in Table \ref{tab:BGLfitlat}.}
   	\label{tab:SMpredlat}
   \end{table}

As mentioned in the introduction, lattice collaborations, like JLQCD and Fermila MILC, are analyzing the form-factors associated with \bdstlnu decays at non-zero recoil. They have shown their preliminary results on this in a couple of conference proceedings  \cite{Aviles-Casco:2019zop,Kaneko:2019vkx}. 
In this section, we do a toy analysis to study the impact of these preliminary lattice results on the \bdstlnu form-factors. In this part of the analysis, we are neither using any experimental inputs, nor inputs from HQET to constrain the form-factors in \bdstlnu and \bdsttaunu. Rather, the obtained predictions are totally dependent on the inputs from lattice and LCSR. The extracted values of the form-factor parameters from these fits and various observable predictions obtained using them can then be compared to those with the experimental inputs. Since the results are preliminary, we have not used them to extract $|V_{cb}|$, but have given a preliminary prediction of $R(D^*)$. 

\begin{figure}[htbp]
	\begin{center}
		\subfigure[]{%
			\label{fig:fplotlcsrlat}
			\includegraphics[width=0.45\textwidth]{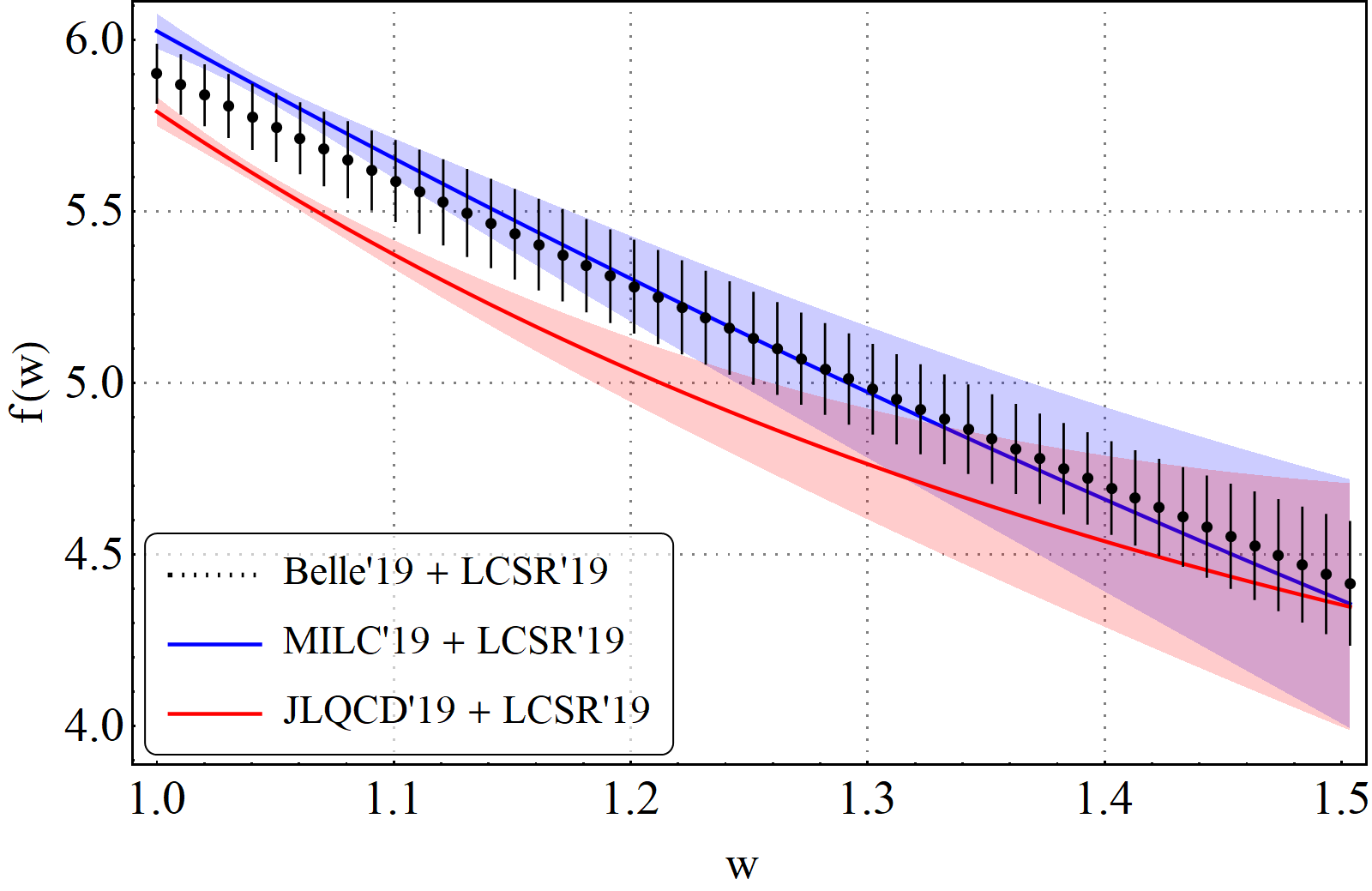}
		}%
		~~~~
		\subfigure[]{%
			\label{fig:f1plotlcsrlat}
			\includegraphics[width=0.45\textwidth]{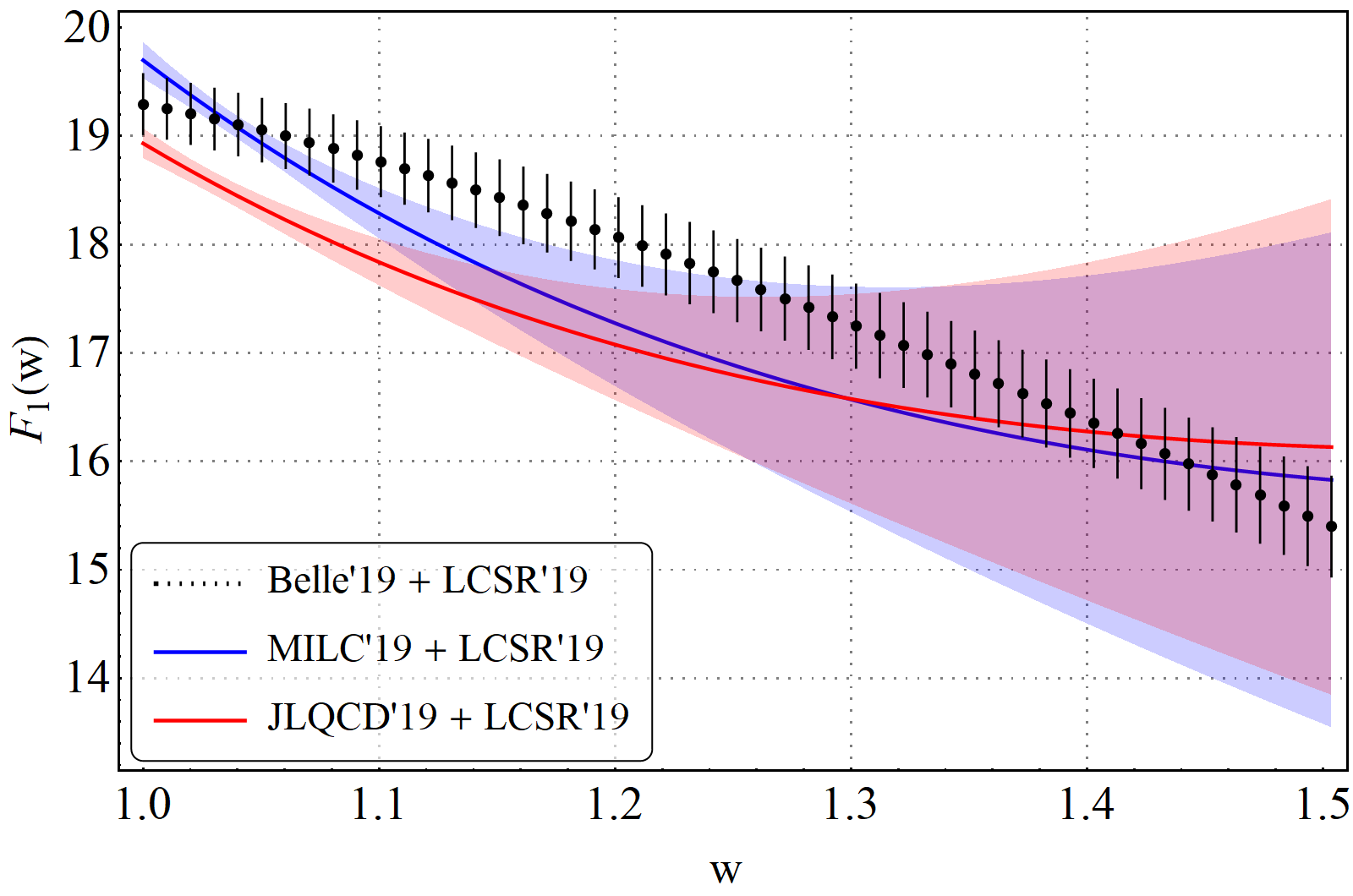}
		}%
		\\
		\subfigure[]{%
			\label{fig:gplotlcsrlat}
			\includegraphics[width=0.453\textwidth]{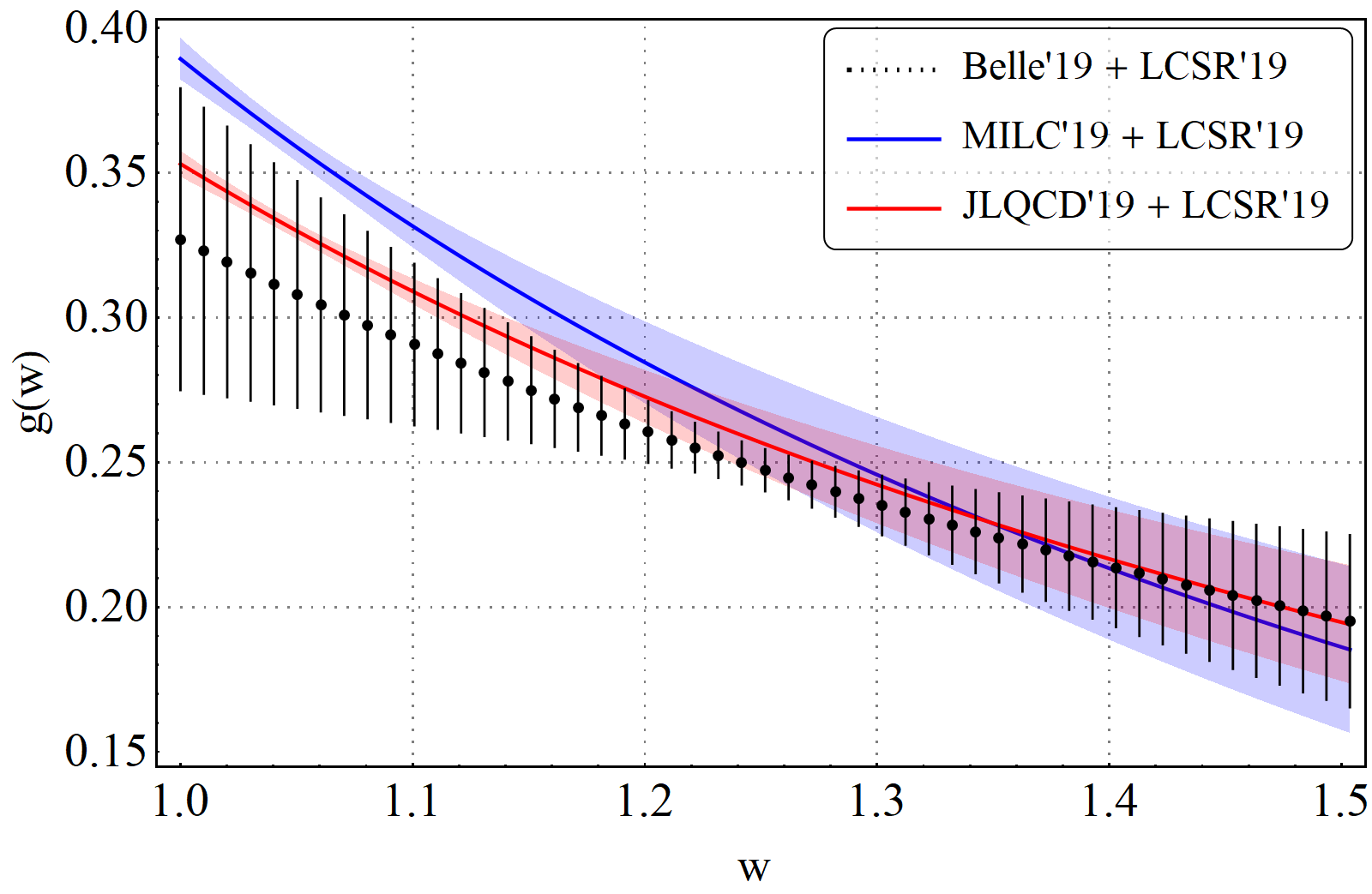}
		}%
		~~~~
		\subfigure[]{%
			\label{fig:f2plotlcsrlat}
			\includegraphics[width=0.451\textwidth]{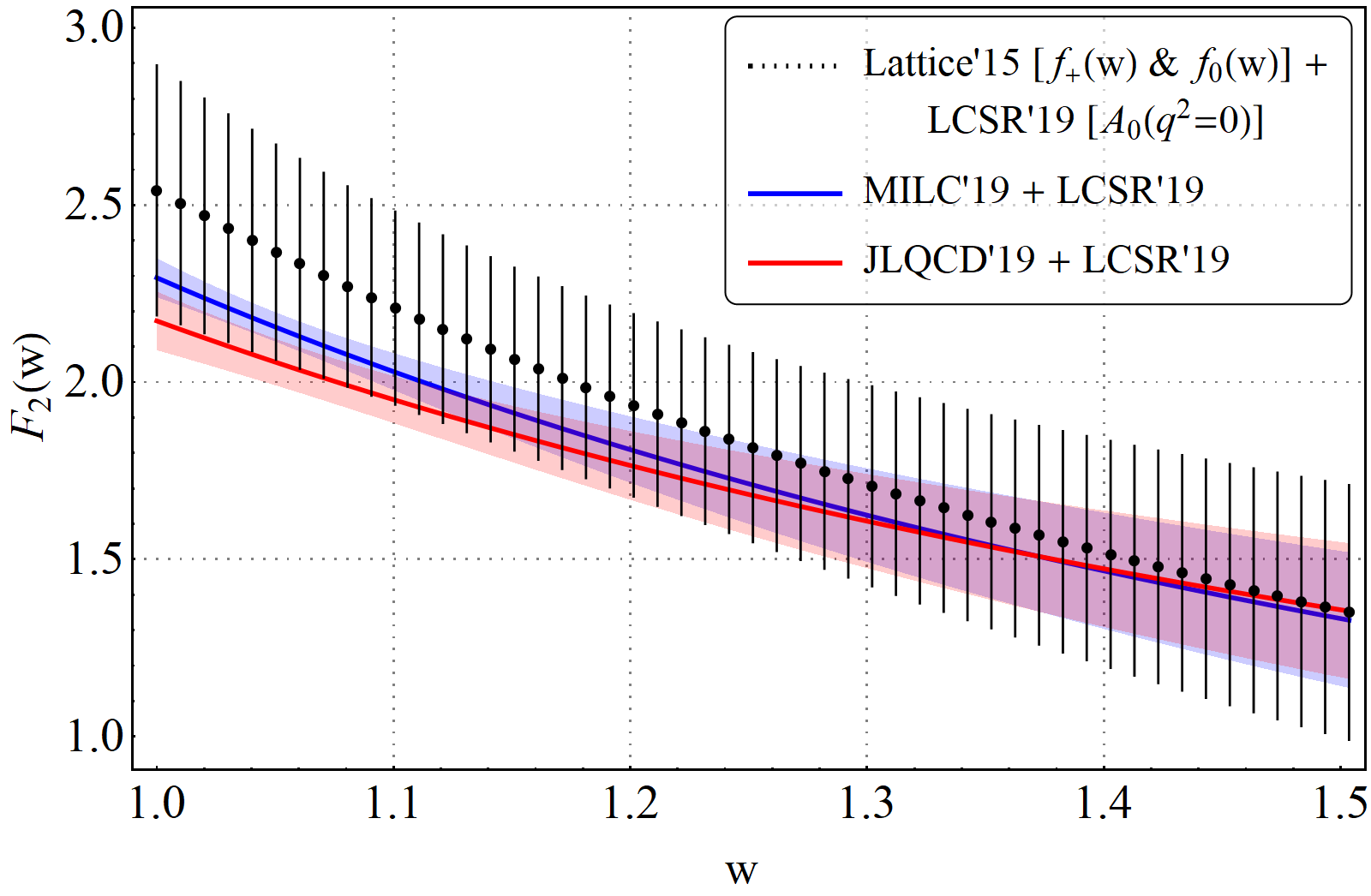}
		}%
	\end{center}
	\caption{In figures \ref{fig:fplotlcsrlat} - \ref{fig:gplotlcsrlat}, the shape of \bdstlnu BGL form factors for $N=2$ obtained from the fit to new Belle data \cite{Abdesselam:2018nnh} (black bars) is compared to that obtained with preliminary lattice inputs from MILC \cite{Aviles-Casco:2019zop} (blue band) and with lattice inputs from JLQCD \cite{Kaneko:2019vkx} (red band). In figure \ref{fig:f2plotlcsrlat}, the blue and red bands hold the same meaning in context to $F_2$ while the black bars represent $F_2$ obtained using the method (involving LCSR inputs\cite{Gubernari:2018wyi}) discussed in section \ref{subsec1b}.}
	\label{fig:ffplotslcsrlat}
\end{figure}

\begin{figure}[t]
	\begin{center}                                                             
		\subfigure[]{%
			\label{fig:bdstlnulat}
			\includegraphics[width=0.47\textwidth]{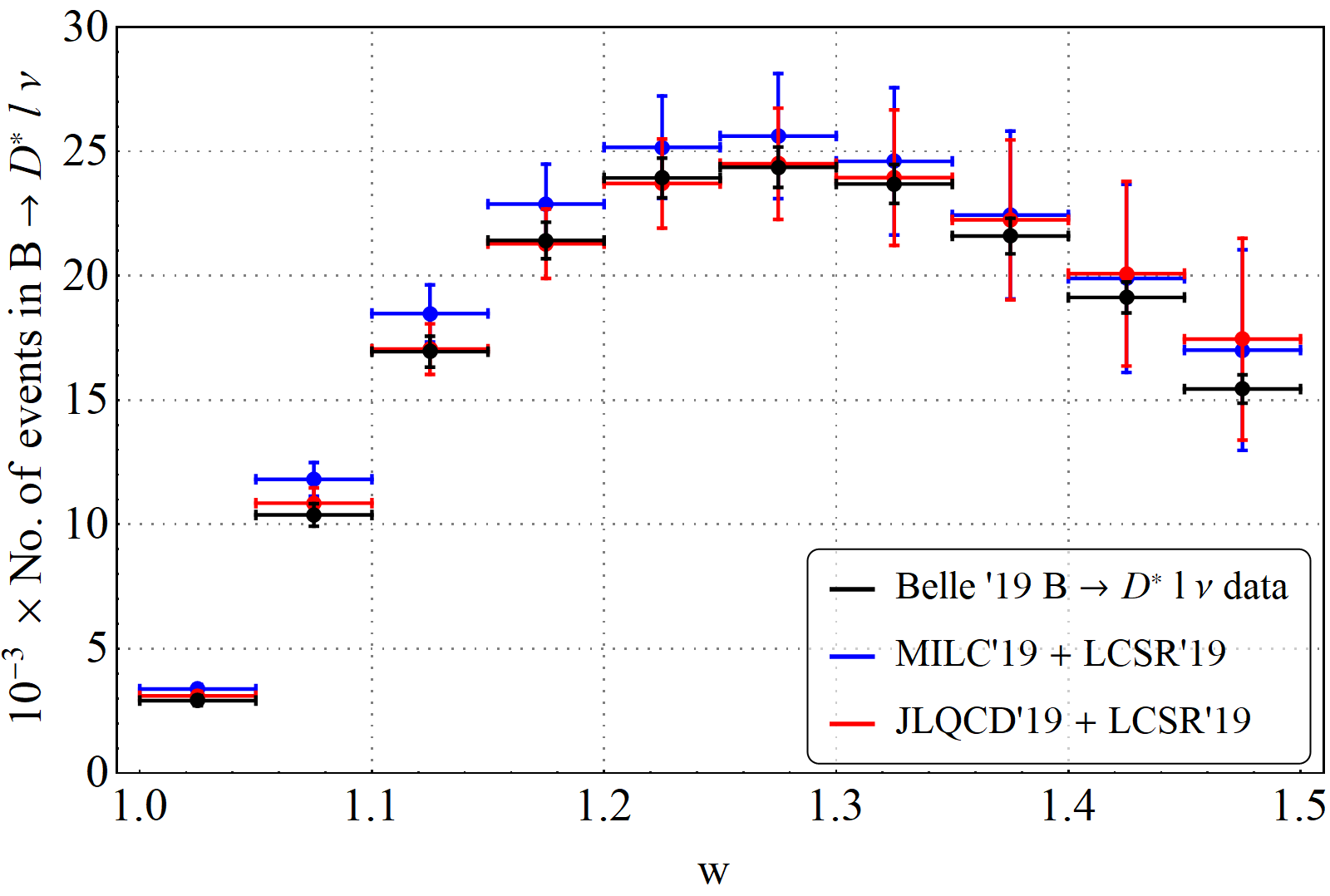}
		}%
		~~~~~
			\subfigure[]{%
				\label{fig:bdstlnulatnorm}
				\includegraphics[width=0.47\textwidth]{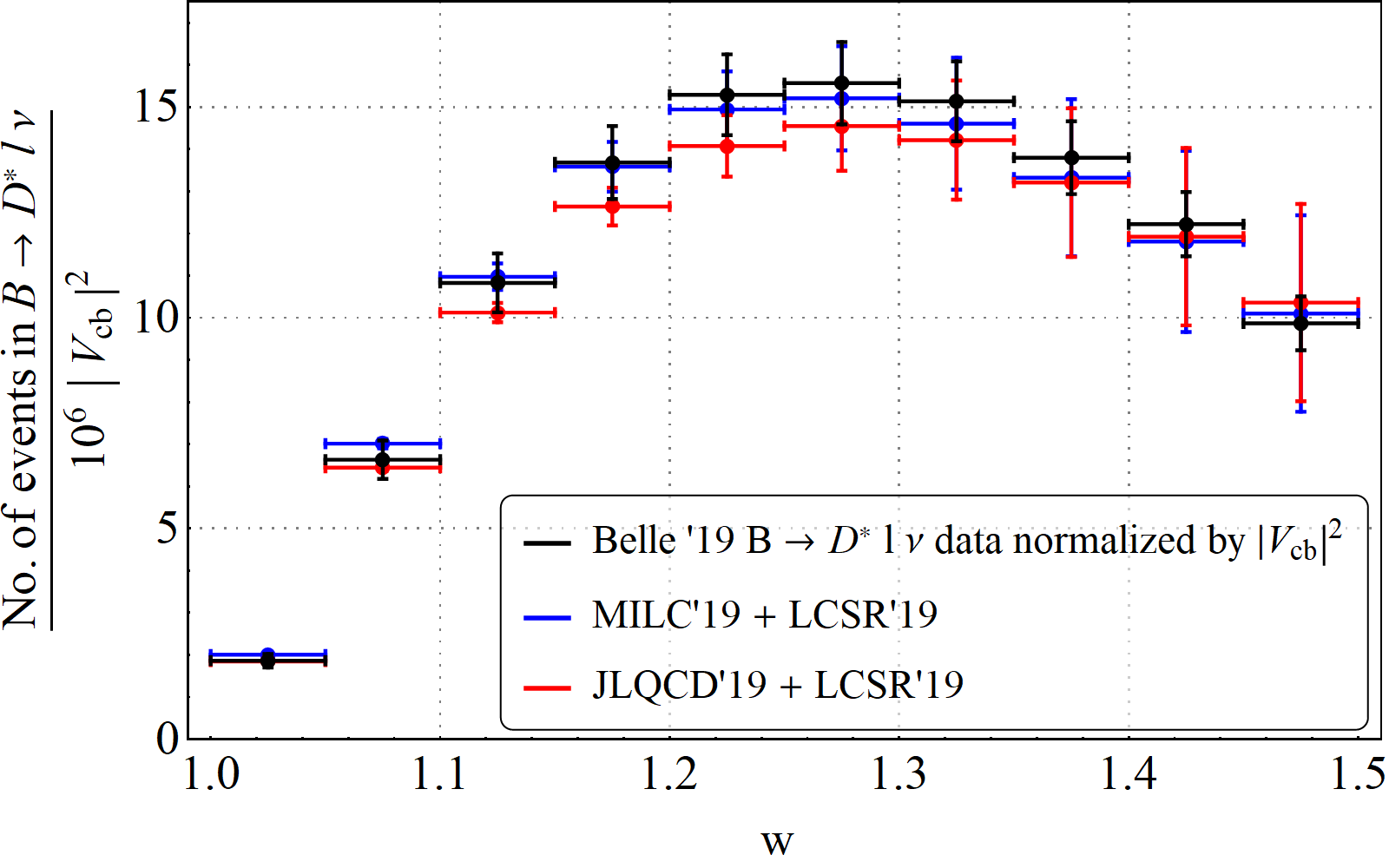}
			}%
\end{center}
\caption{Comparison between the theory and experiment of the \bdstlnu decay distributions in $w$-bins (a) with and (b) without the normalization by $|V_{cb}|^2$. The black error-bars represent the latest experimental data from Belle \cite{Abdesselam:2018nnh}. In the case of theory, the form-factors are extracted using the preliminary inputs from table \ref{tab:BGLfitlat}. }
\label{fig:bdstldis}
\end{figure}
		
\begin{figure}[htbp]
		\begin{center}	
			\includegraphics[width=0.50\textwidth]{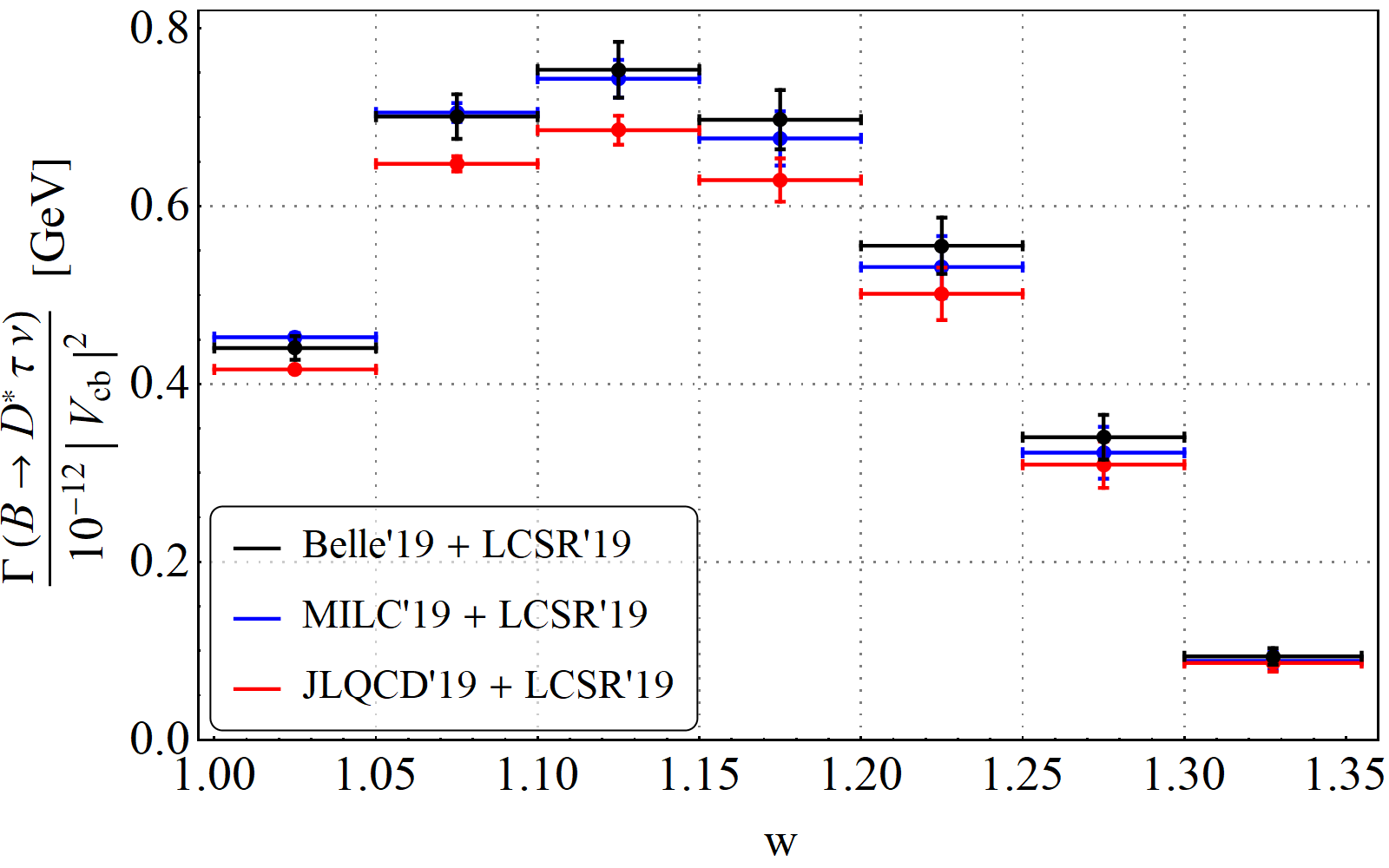}
	\end{center}
	\caption{The differential decay rate distributions (normalized by $|V_{cb}|^2$) in $w$ bins for \bdsttaunu decays in different fit scenarios as explained in the text. The black bars are obtained using our fit results with the Belle 2019 data and LCSR.}
	\label{fig:bdsttaudis}
\end{figure}

In the rest of this section, we discuss the details of our method of analysis. For lattice inputs from JLQCD \cite{Kaneko:2019vkx}, we extract nine lattice data-points for each of the four HQET form factors ($h_V(w), h_{A_1}(w), h_{A_2}(w)$, and $h_{A_3}(w)$) corresponding to nine $w$ values ($w=1, 1.01, 1.02, 1.03, 1.04, 1.05, 1.08, 1.09,$ and $1.10$), i.e., a total of 36 data-points. For the inputs from MILC \cite{Aviles-Casco:2019zop}, we extract five lattice data-points for each of the four HQET form factors corresponding to five $w$ values ($w=1, 1.02, 1.04, 1.06$, and $1.08$), i.e., a total of 20 data-points. The HQET form-factors can be easily expressed in terms of the BGL form-factors ($g(w),~f(w),~F_1(w)$, and $F_2(w)$). Thus, we can easily extract all the required BGL coefficients and form-factors using these preliminary lattice results. Fit results for all four \bdstlnu BGL form-factors for $N=2$ using the JLQCD and MILC data-points are given in the second and third columns of table \ref{tab:BGLfitlat} respectively. In both of these fits, LCSR inputs \cite{Gubernari:2018wyi} for $q^2=0$ are used in addition as well. For $h_{A_1}(1)$, the result from unquenched Fermilab/MILC lattice data \cite{Bailey:2014tva} ($h_{A_1}(1) = 0.906(13)$) is used instead of the preliminary MILC \cite{Aviles-Casco:2019zop} and JLQCD \cite{Kaneko:2019vkx} results. Since we can directly extract $F_2(z)$ using lattice in this case, there is no need to use the HQET relations between form-factors. Using these results, the prediction of $R(D^*)$ and other important observables is straight-forward and they are given in table \ref{tab:SMpredlat} alongwith their correlations. Comparing these results with the ones given in table \ref{tab:SMpred}, we note that all the respective predictions are consistent with each other within the error-bars. Though the central value of $R(D^*)$ obtained from JLQCD results is slightly lower than that found using the other inputs, at the moment, it still has large uncertainty. We have to wait for the final conclusion till we get the final results with complete error budget.   

Figure \ref{fig:ffplotslcsrlat} compares the shape of the form-factors at different values of $w$. The plots are generated using the above mentioned preliminary lattice results at non-zero recoils, and these are compared with the one extracted in our previous analysis (section \ref{sec1}) using experimental results and LCSR. For all the form-factors except $F_2$, the JLQCD and MILC results differ from each other near the zero recoil, which is evident from their data as well. We have used LCSR inputs at $q^2 = 0$, and thus the shape of the form-factors across all the fits agree with each other at the maximum recoil. Considering the uncertainties at 2$\sigma$ CI in the form-factors extracted from experimental data, they agree with the respective JLQCD and MILC predictions then at lower values of recoil. In the extraction of $F_2$, apart from using the preliminary results from JLQCD and MILC, the lattice inputs on $f_+$ and $f_0$ are used together with LCSR (without any experimental inputs). As we can see from figure \ref{fig:f2plotlcsrlat}, the $w$-distributions are agreeing with each other in all the three cases.         

In figure \ref{fig:bdstldis}, we have compared the shape of the differential decay rate distributions in $w$-bins for the \bdstlnu decays for the different fit scenarios as discussed above. Note that while we fit the BGL coefficients using only the new lattice results from JLQCD and MILC, we are not simultaneously extracting $|V_{cb}|$. However, if we want to predict the decay rate distributions, we need to know the value of $|V_{cb}|$ along with all the other relevant inputs. Here, there will always be an ambiguity in the choices of $|V_{cb}|$ while predicting the decay rate distribution. In particular, while we compare these rate distributions with the one obtained from a fit to experimental data. Also, the lattice estimates of the form factors near the zero recoil have tiny errors. Hence, the overall uncertainty in the estimate of the decay rate distribution will be dominated by the uncertainty associated with $|V_{cb}|$. To avoid such circumstances, it will be appropriate to define an observable $\frac{1}{|V_{cb}|^2} \frac{d\Gamma(B \to D^*\tau\nu_{\tau})}{dw}$ where the decay rate distribution is normalised by $|V_{cb}|^2$. 
	
It can be seen from figure \ref{fig:bdstlnulat} that the differential rate distributions predicted using the inputs separately from MILC or JLQCD are in good agreement with the corresponding data from Belle (2019). Also, in the predictions using lattice, we have chosen $|V_{cb}| = 41.04(113)\times 10^{-3}$ which is the value obtained from our analysis of \bdlnu decays in \cite{Jaiswal:2017rve}\footnote{It is important to note that in the extraction of $|V_{cb}|$ from \bdlnu decays, lattice results at zero and non-zero recoils by MILC and HPQCD (table \ref{tab:latinput}) play an important role.}. A similar comparison is shown in figure \ref{fig:bdstlnulatnorm} for the rate distributions normalized by $|V_{cb}|^2$. For normalizing the data, we have used the value of $|V_{cb}| = 39.37\left(\begin{smallmatrix} 107 \\ 121 \end{smallmatrix}\right)$ as given in the table \ref{tab:BGLfitlat}. Here, also the data from Belle is consistent with the results obtained using only the lattice. However, in the first three bins  with low $w$ values, we notice discrepancies between the results obtained using JLQCD and MILC, which were overshadowed in figure \ref{fig:bdstlnulat} by the error in $|V_{cb}|$.      

A similar type of comparison as in figure \ref{fig:bdstlnulatnorm} has been made in figure \ref{fig:bdsttaudis} for \bdsttaunu decays. Here, since we do not have any data in $w$ bins, we have estimated the normalized decay rate distributions using our fit results to Belle 2019 data and LCSR. Similar estimates are done with the fitted results given in table \ref{tab:BGLfitlat}. These predictions will be useful to check in the future experiments like Belle-II. Notice that for high $w$ values, the decay rate distributions in all the three cases are in good agreement with each other, however, in low recoil regions the results from JLQCD and MILC have discrepancies approximately at 3 to 4-$\sigma$ C.I. Also, for all values of $w$, we find good agreement between the distributions obtained from MILC and the data-driven fit. Any further conclusion needs to wait for more complete results from the lattice.

\section{Summary}
In this article, we have reanalyzed the \bdstlnu, and \bdsttaunu decays on the basis of new Belle data and have updated the extracted values of $|V_{cb}|$ and $R(D^*)$ in the SM. We have done the analysis with and without the inputs from LCSR at $q^2=0$. Our new results \emph{without} LCSR are:

\begin{align}
\nn |V_{cb}| &= 39.37 \left(\begin{smallmatrix} 107 \\ 121 \end{smallmatrix}\right)   \times 10^{-3}\\
{\rm and}~~~ R(D^*) &= 0.251 \left(\begin{smallmatrix} 4 \\ 5 \end{smallmatrix}\right) \,,
\end{align}
and \emph{after incorporating} LCSR, we get:
\begin{align}
\nn |V_{cb}| &= 39.56 \left(\begin{smallmatrix} 104 \\ 106 \end{smallmatrix}\right) \times 10^{-3}\\
{\rm and}~~~ R(D^*) &= 0.252 \left(\begin{smallmatrix} 6 \\ 7 \end{smallmatrix}\right) \,.
\end{align}

We note that compared to our 2017 analysis \cite{Jaiswal:2017rve}, the respective uncertainties in both $|V_{cb}|$ and $R(D^*)$ have reduced considerably, and they are consistent with each other within the error-bars. We have also predicted several angular observables associated with the \bdsttaunu decays. The SM predictions for $F_L(D^*)$ is consistent with the respective measurement at 2$\sigma$. Also, the predicted value is lower than the corresponding measured value.   

The above predictions of $R(D^*)$ are not entirely consistent with the measured values. This excess can be explained in a model-independent way by assuming the presence of some new vector, scalar or tensor-type operators. We have worked out the constraints on the NP Wilson coefficients associated with such new operators. The analysis and the resultant parameter spaces of the allowed scenarios show that the data still allow large new physics contributions in these decay modes. We note that the NP scenarios with right-handed scalar quark current are disfavored by the data and the most favored scenario is the one with left-handed vector quark current operator (SM type).  

Very recently, lattice collaborations like Fermilab MILC and JLQCD have presented their preliminary results on the HQET form-factors at non-zero recoil. To do a consistency check, we have predicted $R(D^*)$ using only these lattice results with LCSR $(q^2=0)$. The obtained values are consistent with the one mentioned above. Also, the extracted form-factors and the decay rate distributions in \bdstlnu and \bdsttaunu are compared with the one obtained from the analysis with experimental data on \bdstlnu as inputs.

For all values of $w$, we find agreement in $F_2(w)$ obtained in the three different fit scenarios, however, for the other three form-factors, $F_1(w)$, $f(w)$ and $g(w)$, there are discrepancies in the extracted values in the low recoil regions. In the case of \bdstlnu decays, we find that the available data on the differential $w$-rate distributions are fully consistent with the predictions obtained using preliminary lattice results from MILC and JLQCD, separately. However, in the low recoil regions, in a few bins, the predictions from JLQCD and MILC are not consistent with each other. Similarly, in the case of \bdsttaunu decay, results obtained from Fermilab-MILC agree well with that from fitting the data with lattice and LCSR together. However, in the low recoil regions, both these distributions have discrepancies with the one obtained using inputs from JLQCD. In the large recoil regions, the distributions obtained in all the three different scenarios are in good agreement. More concrete results with appropriate error budgets are needed from the lattice groups for any further conclusion.

\begin{acknowledgments}
	We would like to thank Alejandro Vaquero Avilés-Casco and Paolo Gambino for useful discussions. This project and S.N. is supported by the Science and Engineering Research Board, Govt. of India, under the grant CRG/2018/001260.  
\end{acknowledgments}

\bibliography{sneha_p2}

\end{document}